\documentclass[twocolumn]{aa520}
\usepackage{psfig}
\usepackage{epsfig}
 
\def\kms{$\mathrm{km\;s}^{-1}$} 
 
\def\ha{H$\alpha$}
\def\h2{H$_{2}$}

\def\niipg{[N~{\small II}]$\,\lambda\lambda6548,6583$}

\def\siipg{[S~{\small II}]$\,\lambda\lambda6716,6731$}

\def\kmsmpc{$\rm km\;s^{-1}\;Mpc^{-1}$} 
\def\vs{$v_\star$}
\def\Vs{$V_\star$}
\def\vg{$v_g$}
\def\Vg{$V_g$}
\def\ss{$\sigma_\star$}
\def\sg{$\sigma_g$}
\def\h3{$h_{3}$}
\def\h4{$h_{4}$}
\def\FWHM{{\it FWHM\/}}
\def\ap{{\approx\,}}
  
\begin{document} 
 
\title{Minor-axis velocity gradients in spirals and the case of inner
  polar disks \thanks{Based on observations carried out at the
  European Southern Observatory (ESO 62.A-0463 and
  63.N-0305).}$^{\bf,}$\thanks{Tables 3 to 4 are only available in
  electronic form at the CDS via anonymous ftp to cdsarc.u-strasbg.fr
  (130.79.128.5) or via http://cdsweb.u-strasbg.fr/Abstract.html.}}

\author{ 
     E.M.~Corsini,           
     A.~Pizzella,            
     L.~Coccato,             
     and F.~Bertola}         
 
\offprints{Enrico Maria Corsini}
  
\institute{Dipartimento di Astronomia, Universit\`a di Padova, vicolo
dell'Osservatorio~2, I-35122 Padova, Italy}

\date{Received 4 March 2003; accepted 3 June 2003}

\titlerunning{Minor-axis velocity gradients in spirals} 
\authorrunning{Corsini et al.}

\abstract{
We measured the ionized-gas and stellar kinematics along the major and
minor axis of a sample of 10 early-type spirals. Much to our surprise
we found a remarkable gas velocity gradient along the minor axis of 8
of them.
According to the kinematic features observed in their ionized-gas
velocity fields, we divide our sample galaxies in three classes of
objects. (i) NGC 4984, NGC 7213, and NGC 7377 show an overall velocity
curve along the minor axis without zero-velocity points, out to the
last measured radius, which is interpreted as due to the warped
structure of the gaseous disk. (ii) NGC 3885, NGC 4224, and NGC 4586
are characterized by a velocity gradient along both major and minor
axis, although non-zero velocities along the minor axis are confined
to the central regions. Such gas kinematics have been explained as
being due to non-circular motions induced by a triaxial
potential. (iii) NGC 2855 and NGC 7049 show a change of slope of the
velocity gradient measured along the major axis (which is shallower in
the center and steeper away from the nucleus), as well as non-zero gas
velocities in the central regions of the minor axis. This has been
attributed to the presence of a kinematically-decoupled gaseous
component in orthogonal rotation with respect to the galaxy disk,
namely an inner polar disk.  The case and origin of inner polar disks
are discussed and the list of their host galaxies is presented.
}

\maketitle

\section{Introduction} 
\label{sec:introduction} 

As a result of the analysis of the kinematical data available in the
literature for the ionized-gas component of the S0s and spirals listed
in the Revised Shapley-Ames Catalog of Bright Galaxies (Sandage \&
Tammann 1981, hereafter RSA), we realized that $\sim60\%$ of unbarred
galaxies show a remarkable gas velocity gradient along their optical
minor axis. This phenomenon is observed all along the Hubble sequence
of disk galaxies, but it is particularly frequent in early-type
spirals (Coccato et al. 2003). However, such a minor-axis velocity
gradient is unexpected if gas is moving onto circular orbits in a disk
coplanar to the stellar one.

In a non-axisymmetric potential, gas in equilibrium moves onto closed
elliptical orbits which become nearly circular at larger radii (de
Zeeuw \& Franx 1989; Gerhard, Vietri \& Kent 1989).  A velocity
gradient along the apparent minor axis of the galaxy is observed if
the inner gas orbits are seen at intermediate angle between their
intrinsic major and minor axes.
This is the case of bulges, since their intrinsic shape is generally
triaxial both in barred (Kormendy 1982) and in unbarred galaxies
(Bertola, Vietri \& Zeilinger 1991). Detailed velocity fields
for the gaseous component have been obtained and modeled to
demonstrate bulge triaxiality only for our Galaxy (Gerhard \& Vietri
1986), and two external galaxies, namely NGC 4845 (Bertola, Rubin \&
Zeilinger 1989; Gerhard et al. 1989) and M31 (Berman 2001 and
references therein).

A gas velocity gradient along the apparent minor axis of the galaxy is
observed if the gaseous component is not settled onto the galaxy disk,
as in the case of kinematically-decoupled gaseous components and
warped gaseous disks.
The presence of a kinematically-decoupled component is usually
interpreted as the relic of accretion events that occurred during the
host galaxy lifetime. The nature and the orientation of the decoupled
component mostly depend on the initial angular momentum and mass of
the accreted material.
Polar rings (e.g. Whitmore et al. 1990), counterrotating disks (see
Galletta 1996; Bertola \& Corsini 1999 for a review) and
orthogonally-rotating cores (e.g. Bertola \& Corsini 2000; Sil'chenko
2003) are examples of kinematically-decoupled structures observed in
disk galaxies, most of which display an otherwise undisturbed
morphology.

\begin{table*}[!ht]
\caption{Parameters of the sample galaxies} 
\begin{tabular}{lllrrccrrc}
\hline 
\noalign{\smallskip} 
\multicolumn{1}{c}{Name} & 
\multicolumn{2}{c}{Morphological type} &
\multicolumn{1}{c}{P.A.} & 
\multicolumn{1}{c}{$V_\odot$} & 
\multicolumn{1}{c}{$D$} & 
\multicolumn{1}{c}{$M^0_{B_T}$} & 
\multicolumn{1}{c}{$D_{25} \times d_{25}$} &
\multicolumn{1}{c}{$i$} &  
\multicolumn{1}{c}{Minor-axis region} \\
\noalign{\smallskip}
\multicolumn{1}{c}{} &  
\multicolumn{1}{c}{[RSA]} & 
\multicolumn{1}{c}{[RC3]} & 
\multicolumn{1}{c}{[$^\circ$]} &
\multicolumn{1}{c}{[km s$^{-1}$]} & 
\multicolumn{1}{c}{[Mpc]} &
\multicolumn{1}{c}{[mag]} & 
\multicolumn{1}{c}{} & 
\multicolumn{1}{c}{[$^\circ$]} &
\multicolumn{1}{c}{where $V_g\neq0$} \\
\noalign{\smallskip}
\multicolumn{1}{c}{(1)} &
\multicolumn{1}{c}{(2)} & 
\multicolumn{1}{c}{(3)} & 
\multicolumn{1}{c}{(4)} & 
\multicolumn{1}{c}{(5)} & 
\multicolumn{1}{c}{(6)} & 
\multicolumn{1}{c}{(7)} & 
\multicolumn{1}{c}{(8)} &
\multicolumn{1}{c}{(9)} & 
\multicolumn{1}{c}{(10)}\\ 
\noalign{\smallskip} 
\hline 
\noalign{\smallskip} 
NGC~1638& Sa                & SAB0(rs)?    &  70& $3258\pm11$& 43.5& $-20.42$& $2\farcm0\times1\farcm5$& 43& ?\\ 
NGC~2855& Sa(r)             & (R)S0/a(rs)  & 120& $1897\pm17$& 22.3& $-19.45$& $2\farcm5\times2\farcm2$& 27& inner\\
NGC~3885& Sa                & S0/a(s)      & 123& $1952\pm 8$& 22.3& $-19.44$& $2\farcm4\times1\farcm0$& 69& inner\\
NGC~4224& Sa                & Sa(s): sp    &  57& $2624\pm 6$& 33.0& $-20.07$& $2\farcm6\times1\farcm0$& 69& inner\\
NGC~4235& Sa                & Sa(s) sp     &  48& $2249\pm11$& 28.1& $-20.35$& $4\farcm2\times0\farcm9$& 82& ?\\
NGC~4586& Sa                & Sa(s): sp    & 115&  $815\pm 7$& 17.0& $-19.04$& $4\farcm0\times1\farcm3$& 73& inner\\
NGC~4984& Sa(s)             & (R)SAB0$^{+}$&  90& $1279\pm 6$& 14.2& $-18.73$& $2\farcm8\times2\farcm2$& 38& extended\\
NGC~7049& S0$_{3}$(4)/Sa    & S0(s)        &  57& $2285\pm 6$& 29.6& $-20.79$& $4\farcm3\times3\farcm0$& 47& inner\\
NGC~7213& Sa(rs)            & Sa(s):       & 124& $1784\pm 6$& 23.1& $-20.69$& $3\farcm1\times2\farcm8$& 27& extended\\
NGC~7377& S0$_{2/3}$/Sa pec & S0$^{+}$(s)  & 101& $3339\pm11$& 45.7& $-21.37$& $3\farcm0\times2\farcm5$& 34& extended\\ 
\noalign{\smallskip}  
\hline 
\noalign{\smallskip}  
\end{tabular}\\  
\begin{footnotesize}
\begin{minipage}{18cm} 
NOTES. -- 
Col. 2: Morphological classification from RSA.
Col. 3: Morphological classification from RC3.
Col. 4: Major-axis position angle from RC3. 
Col. 5: Heliocentric systemic velocity derived from the center of 
        symmetry of the rotation curve of the gas along the
        galaxy major axis.
Col. 6: Distance obtained as $V_0/H_0$ with $H_0=75$ \kmsmpc\ 
        and $V_0$ the systemic velocity derived from $V_\odot$ 
        corrected for the motion of the Sun with respect of the 
        Local Group according to the RSA. For NGC 4586 we assumed 
        the distance of the Virgo Cluster by Freedman et al. (1994).
Col. 7: Absolute total blue magnitude corrected for 
        inclination and extinction from RC3.
Col. 8: Major and minor isophotal diameters from RC3.
Col. 9: Inclination derived from $\cos^{2}{i}\,=\,(q^2-q_0^2)/(1-q_0^2)$ 
        assuming the observed axial ratio $q=d_{25}/D_{25}$ from RC3
        and the typical intrinsic axial ratio for S0/Sa and Sa galaxies 
        ($q_0=0.18$, Guthrie 1992).  
Col. 10: Minor-axis region where non-zero velocity is measured for
         the ionized gas. Inner = non-zero velocity is 
         confined to the nuclear region; extended = non-zero velocity is
         observed out to the last measured radius; ? = doubtful velocity 
         gradient.
\end{minipage} 
\end{footnotesize}
\label{tab:sample}
\end{table*}

A misalignment between the angular momenta of inner and outer parts
characterizes a warped disk. This phenomenon has been observed in at
least half of all the spiral galaxies and has been explained both with
an internal and external origin (see Binney 1992 for a review).
Significant warps are usually measured in the cold gaseous component,
while they are less frequent in the stellar one. Gaseous and stellar
disks are generally coplanar and flat in the optical region and their
warp is observed at larger radii (Briggs 1990; Bosma 1991), though
non-coplanar gas and stellar disks have been seen even in the
optical region of some galaxies giving rise to a minor-axis velocity
gradient.
For example, in a triaxial potential bulge (or in a bar) which is
tumbling about the short axis, the gas leaves the plane perpendicular
to the rotation axis and settles onto the so-called anomalous orbits
(van Albada, Kotanyi \& Schwarzschild 1982). These closed orbits are
fixed in the figure even though the figure rotates. They are nearly
planar and fairly circular, but their inclination with respect to the
figure rotation axis increases with the orbital radius.  This explains
the tilted disks of counterrotating gas observed in some SB0 galaxies
(e.g. Emsellem \& Arsenault 1996).

This paper aims to investigate the presence of non-circular and
off-plane gas motions in bulges. These phenomena have to be taken into
account as well as the presence of pressure-supported ionized gas
(Bertola et al. 1995; Cinzano et al. 1999; Pignatelli et al. 2001)
when the observed rotation velocity of the gas is adopted to trace the
circular speed in mass modeling, in order to derive the distribution
of luminous and dark matter in the innermost regions of disk galaxies.

\begin{figure*} 
\begin{minipage}[t]{4.2cm} 
\vspace{0pt}  
\psfig{figure=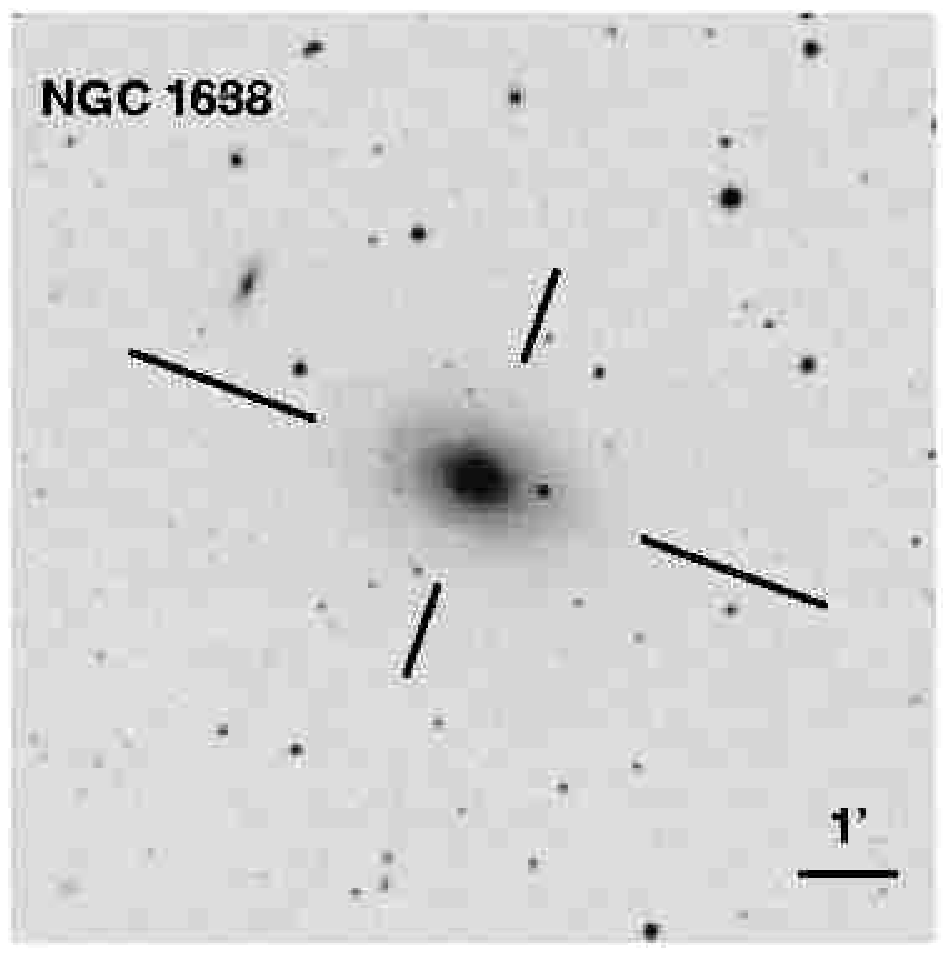,width=4.2cm}
\end{minipage}
\hspace*{0.2cm}
\begin{minipage}[t]{4.2cm}
\vspace{0pt}
\psfig{figure=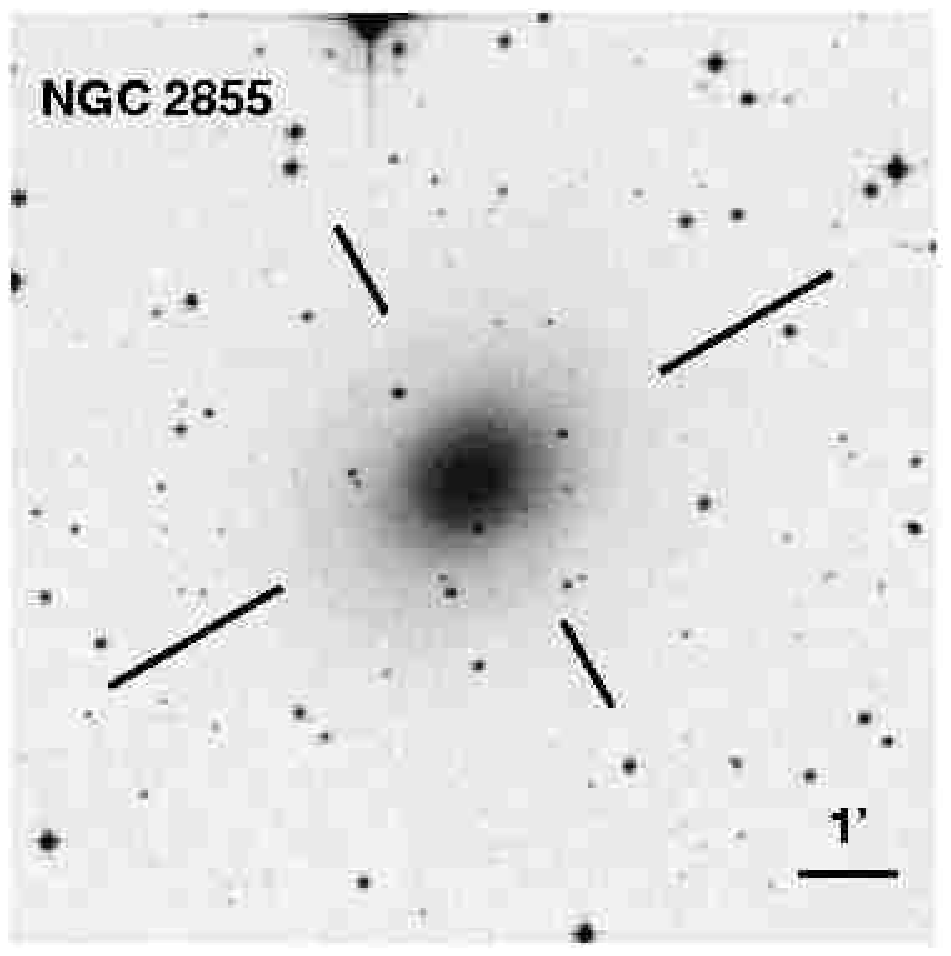,width=4.2cm}
\end{minipage}
\hspace*{0.2cm}
\begin{minipage}[t]{4.2cm}
\vspace{0pt}
\psfig{figure=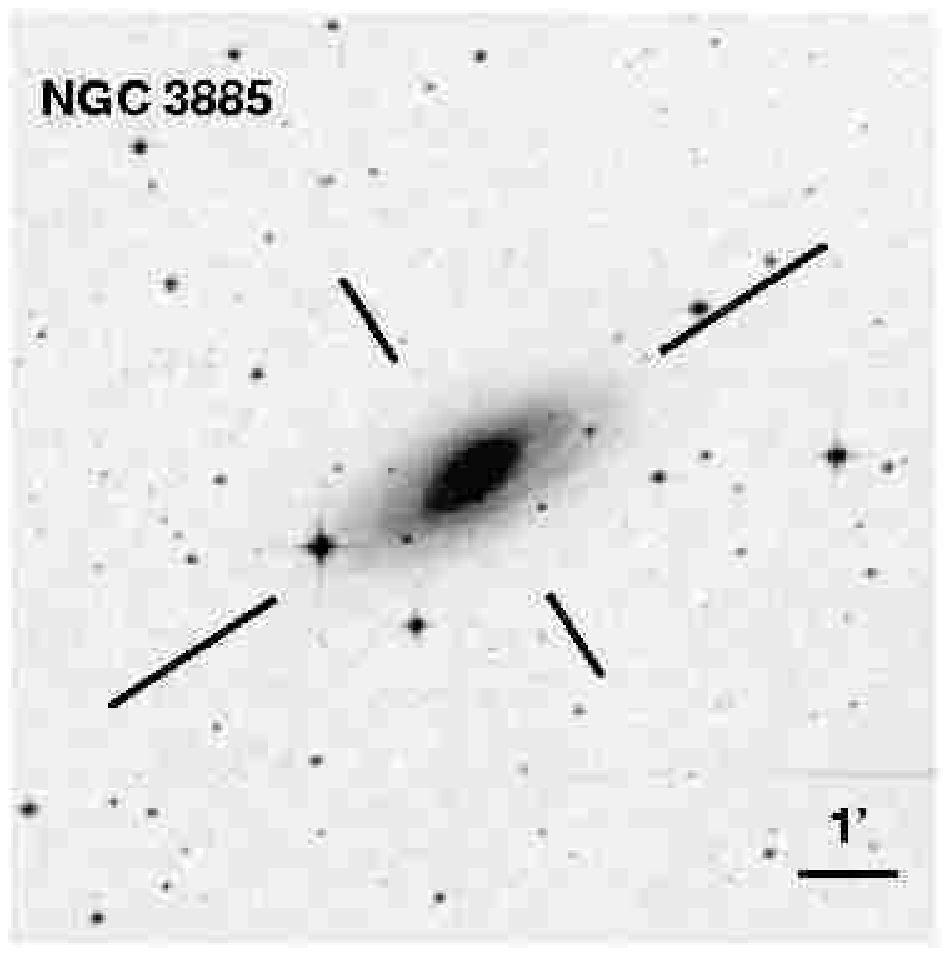,width=4.2cm}
\end{minipage}
\hspace*{0.2cm}
\begin{minipage}[t]{4.2cm}
\vspace{0pt}
\psfig{figure=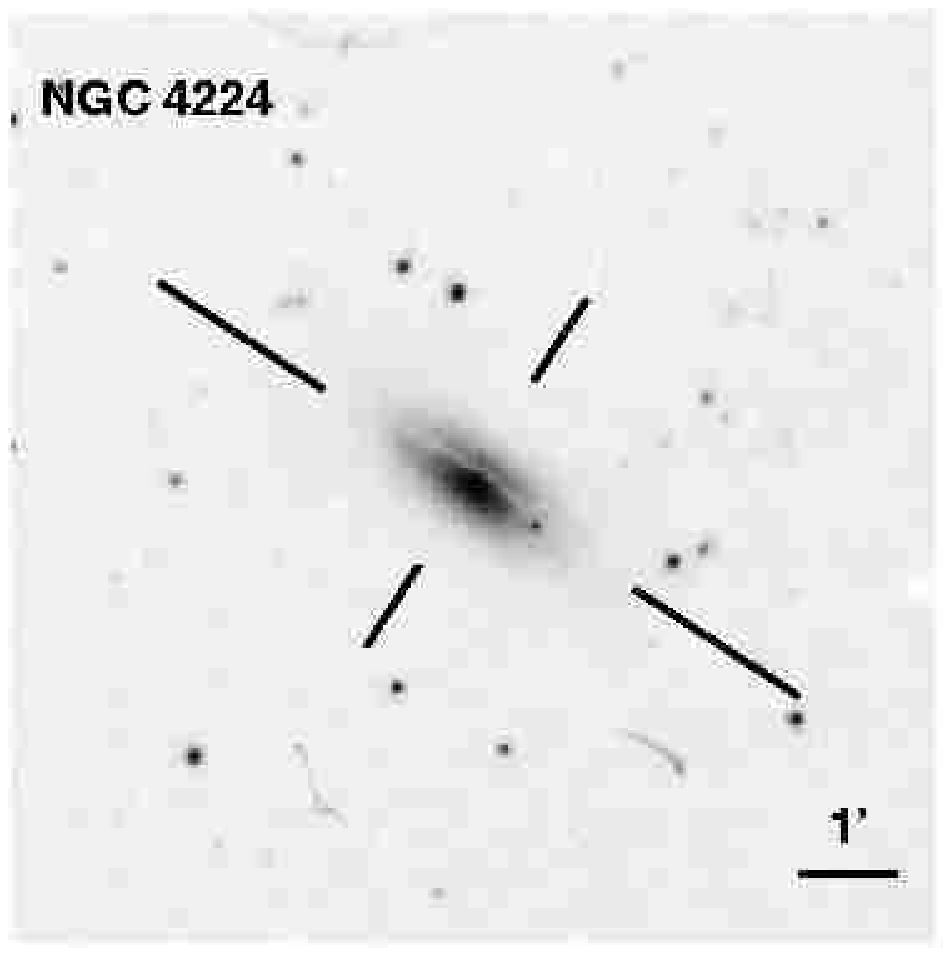,width=4.2cm}
\end{minipage}

\begin{minipage}[t]{4.2cm} 
\vspace{10pt}  
\psfig{figure=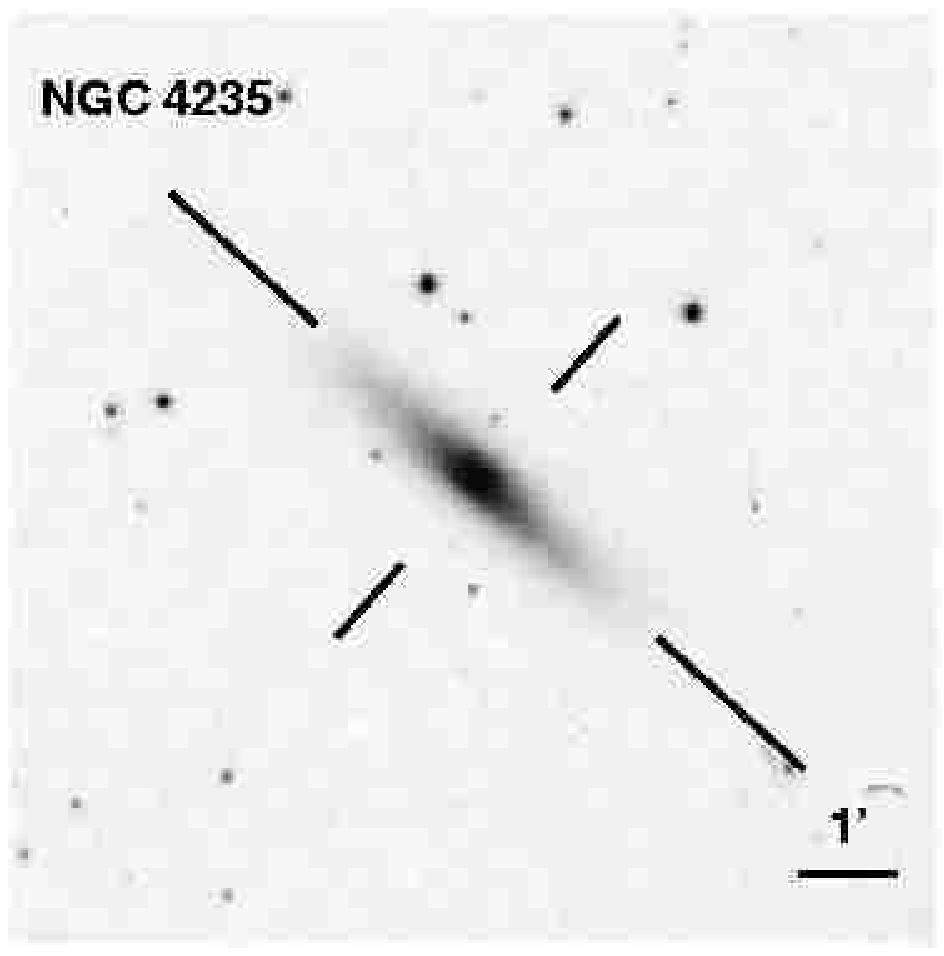,width=4.2cm}
\end{minipage}
\hspace*{0.2cm}
\begin{minipage}[t]{4.2cm}
\vspace{10pt}
\psfig{figure=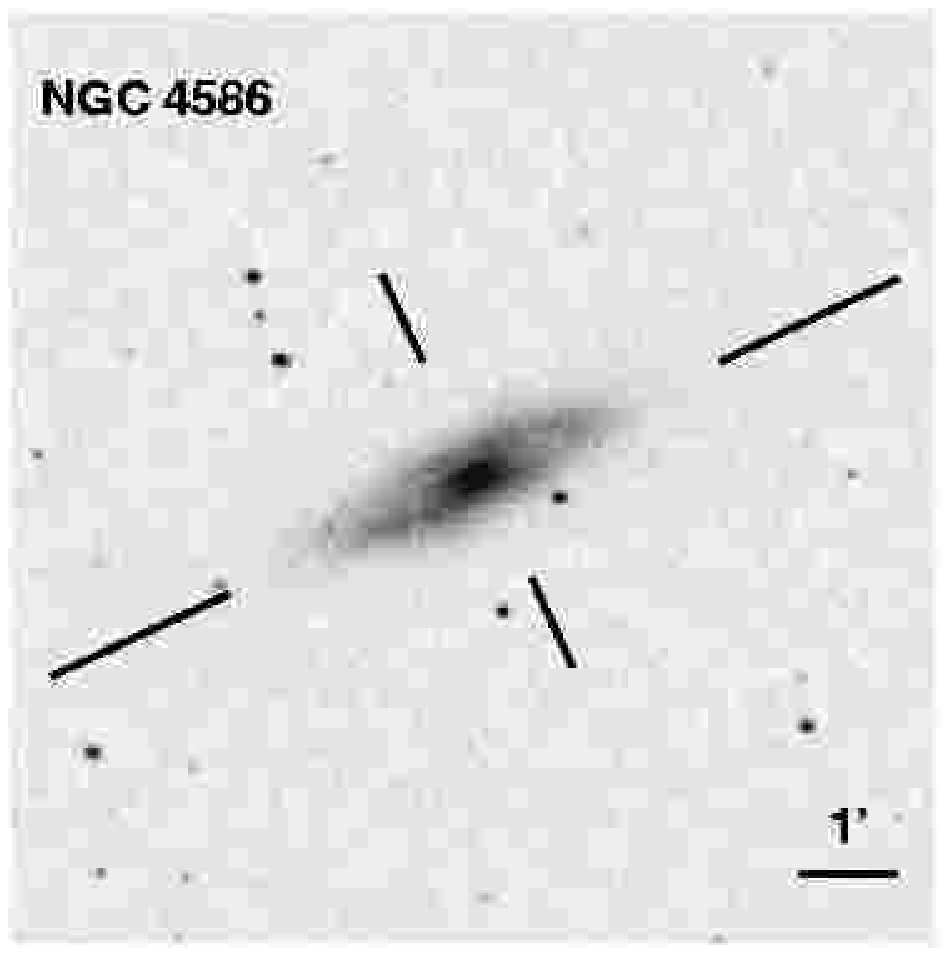,width=4.2cm}
\end{minipage}
\hspace*{0.2cm}
\begin{minipage}[t]{4.2cm}
\vspace{10pt}
\psfig{figure=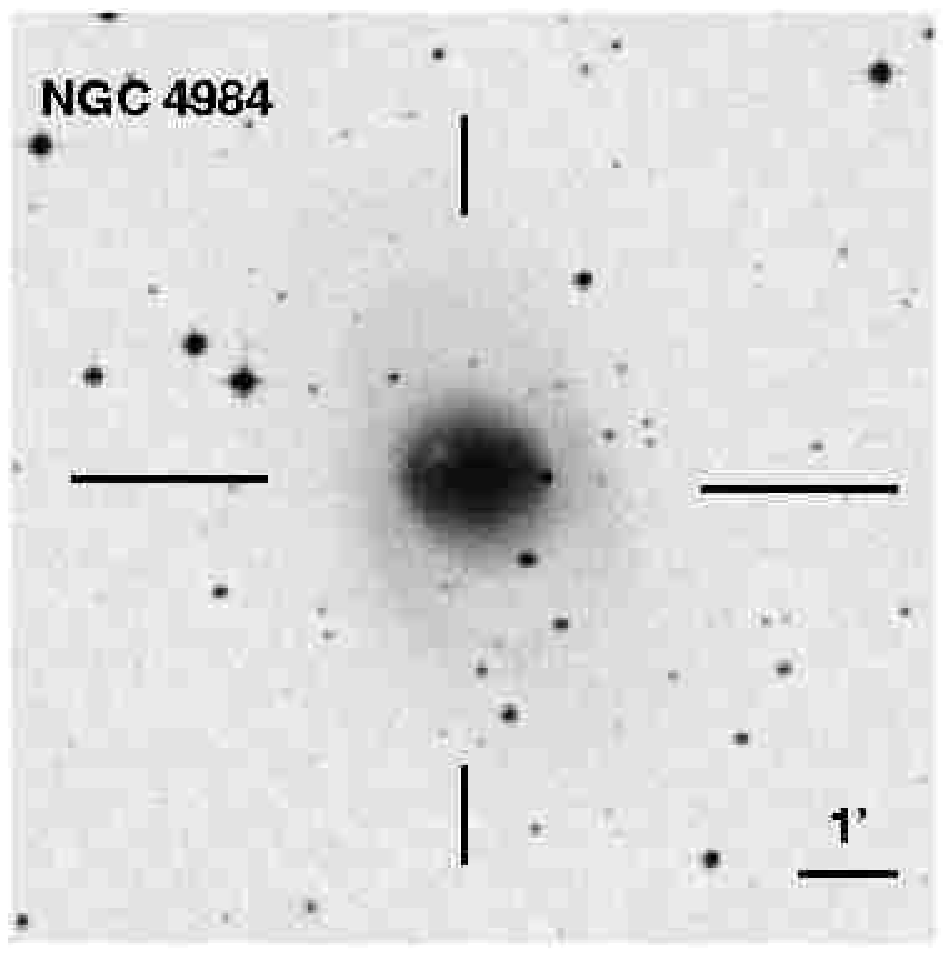,width=4.2cm}
\end{minipage}
\hspace*{0.2cm}
\begin{minipage}[t]{4.2cm}
\vspace{10pt}
\psfig{figure=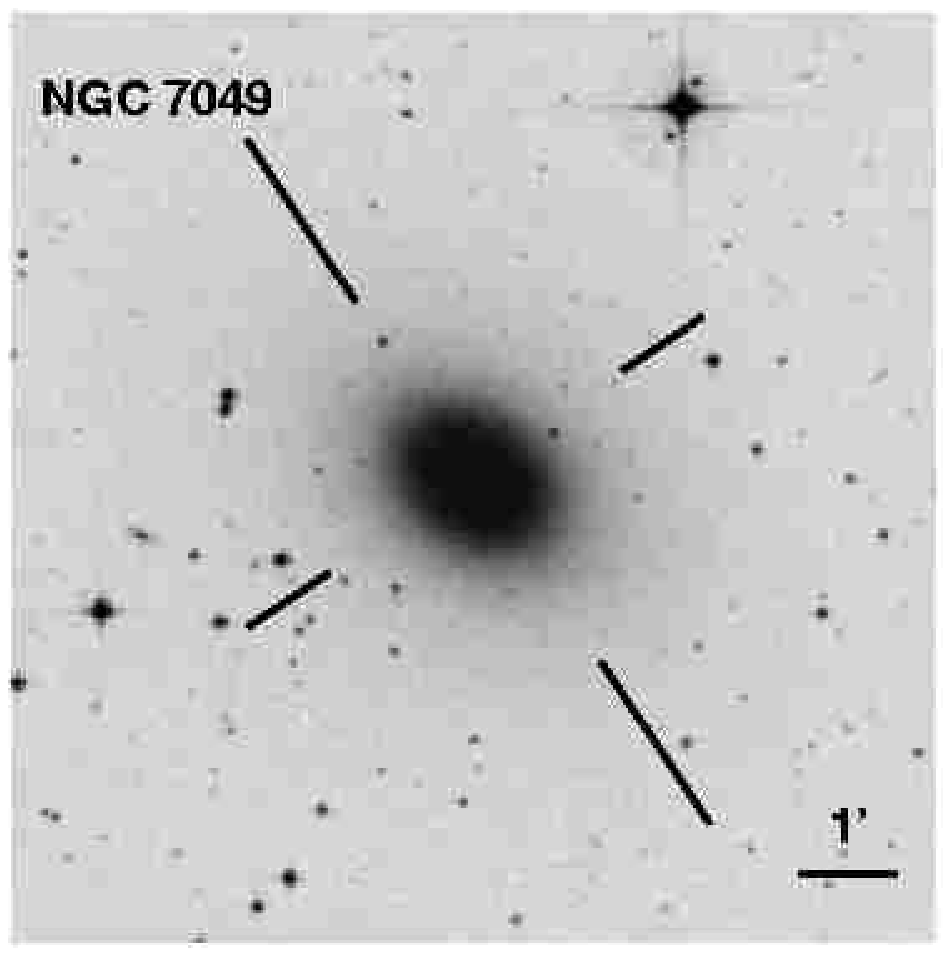,width=4.2cm}
\end{minipage}
\begin{minipage}[t]{4.2cm} 
\vspace{10pt}  
\psfig{figure=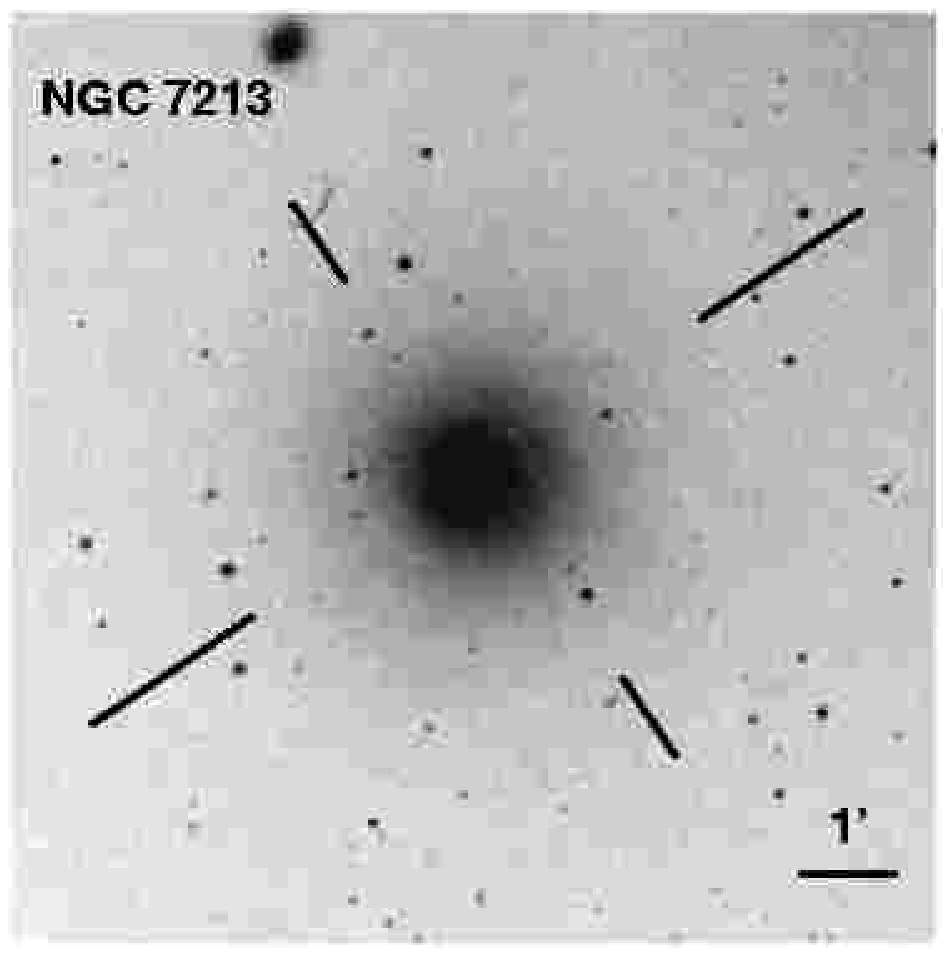,width=4.2cm}
\end{minipage}
\hspace*{0.2cm}
\begin{minipage}[t]{4.2cm}
\vspace{10pt}
\psfig{figure=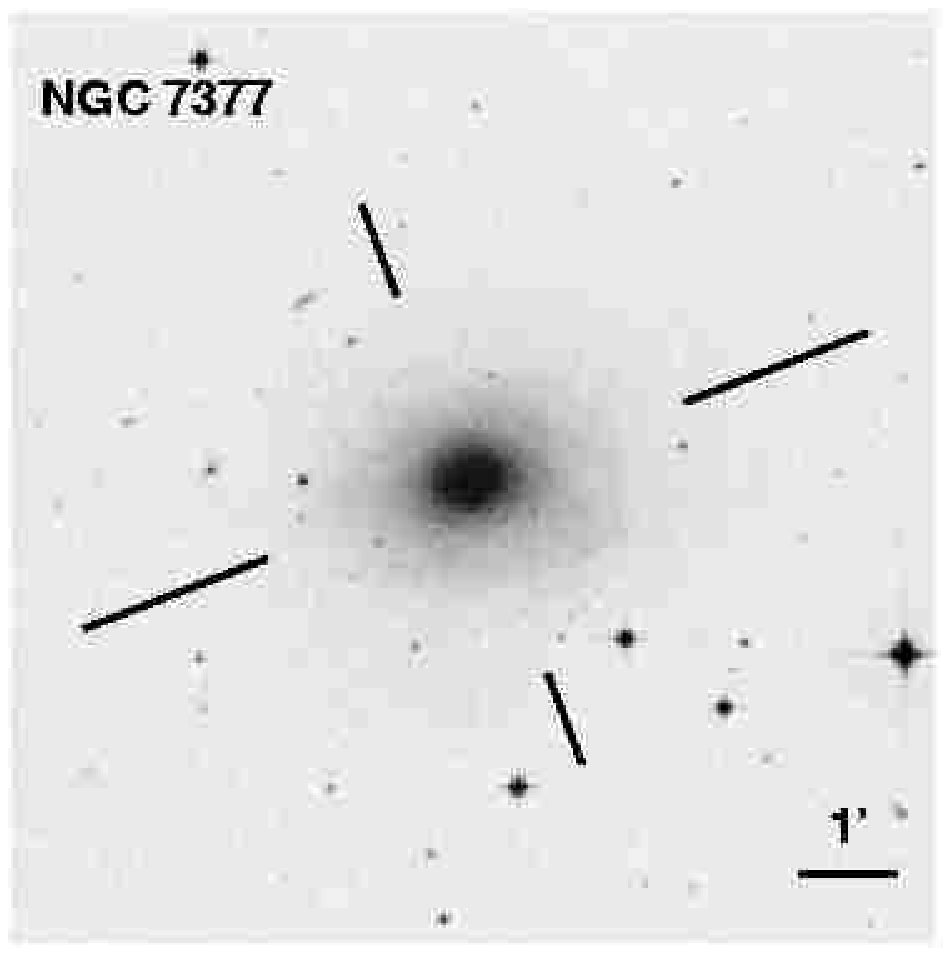,width=4.2cm}
\end{minipage}
\caption{Optical images of the sample galaxies taken from the 
  Digitized Sky Survey. The orientation of the images is north up and
  east left and each image is $8'\times8'$. Long and short lines mark
  the position of optical major and minor axis according to RC3
  catalog.} 
\label{fig:sample}
\end{figure*}

To maximize the chance of detecting a gas velocity gradient along the
minor axis we selected a sample of early-type spirals according to the
results of Coccato et al. (2003).  We randomly chose 10 galaxies among
the Sa spirals listed in The Carnegie Atlas of Galaxies (Sandage \&
Bedke 1994, hereafter CAG). A compilation from the literature of the
properties of the sample galaxies is presented in Table 1. The images
of the sample galaxies taken from the Digitized Sky Survey are given
in Fig.  \ref{fig:sample}. In a previous paper (Corsini, Pizzella \&
Bertola 2002) we reported on the special case of NGC 2855, which is
characterized by the presence of a kinematically-decoupled component
of ionized gas in orthogonal rotation with respect to the stellar disk
as well as of a faint ring-like structure surrounding the
galaxy. These features are indicative of an on-going acquisition
process.  Here we present results for the remaining 9 sample galaxies.
The long-slit spectroscopic observations and data analysis are
described in Sect. \ref{sec:observations}. The resulting stellar and
ionized-gas kinematics are given in Sect. \ref{sec:kinematics} and
interpreted in Sect.  \ref{sec:discussion}. The case of inner polar
disks is discussed in Sect.  \ref{sec:discussion} too.

\section{Spectroscopic observations and data reduction}
\label{sec:observations}

\subsection{Spectroscopic observations}

The spectroscopic observations were carried out at the European
Southern Observatory in La Silla (Chile) with the 1.52-m ESO telescope
on March 17-22 (hereafter run 1) and September 09-12, 1999 (hereafter
run 2).
The telescope was equipped with the Boller \& Chivens Spectrograph.
The No.~33 grating with 1200 grooves mm$^{-1}$ was used in the first
order in combination with a $2\farcs2\times4\farcm2$ slit, and the No.
39 Loral/Lesser CCD with $2048\times2048$ pixels of $15\times15$
$\mu$m$^2$. The spectral range between about 4850 \AA\ and 6850
\AA\ was covered with a reciprocal dispersion of 0.98 \AA\ 
pixel$^{-1}$. The spatial scale was $0\farcs81$ pixel$^{-1}$.  The
instrumental resolution was 2.75 \AA\ (\FWHM) and it was measured from
the mean \FWHM\ of arc lines in wavelength-calibrated comparison
spectra. It corresponds to $\sigma_{\it instr}\approx50$ \kms\ at \ha.
Repeated exposures (typically of 3600 s each) were taken along both
major and minor axes of all the sample galaxies, after centering their
nucleus on the slit using the guiding TV camera. The observed position
angles and corresponding exposure times are given in Table
\ref{tab:log}. In each observing run we obtained spectra of late-G to
early-K giant stars to be used as templates in measuring the stellar
kinematics. A spectrum of the comparison helium-argon lamp was taken
before and/or after every object exposure to ensure an accurate
wavelength calibration.  Seeing \FWHM\ during the observing nights
ranged between $1''$ and $3''$ as measured by the ESO Differential
Image Motion Monitor.

\begin{table*}[ht!] 
\caption{Log of spectroscopic observations} 
\begin{flushleft}  
\begin{tabular}{lcrcrcc} 
\hline 
\noalign{\smallskip} 
\multicolumn{1}{c}{Object} &
\multicolumn{1}{c}{Run} & 
\multicolumn{1}{c}{P.A.} & 
\multicolumn{1}{c}{Position} & 
\multicolumn{1}{c}{Single exp. time} & 
\multicolumn{1}{c}{Total exp. time} & 
\multicolumn{1}{c}{Template}\\ 
\noalign{\smallskip} 
\multicolumn{1}{c}{} &
\multicolumn{1}{c}{} & 
\multicolumn{1}{c}{[$\circ$]}& 
\multicolumn{1}{c}{} & 
\multicolumn{1}{c}{[s]} & 
\multicolumn{1}{c}{[h]} & 
\multicolumn{1}{c}{}\\ 
\noalign{\smallskip}
\multicolumn{1}{c}{(1)} &
\multicolumn{1}{c}{(2)} & 
\multicolumn{1}{c}{(3)} & 
\multicolumn{1}{c}{(4)} & 
\multicolumn{1}{c}{(5)} & 
\multicolumn{1}{c}{(6)} & 
\multicolumn{1}{c}{(7)}\\ 
\noalign{\smallskip} 
\hline 
\noalign{\smallskip} 
NGC~1638 & 2 &  70 & MJ &          3600$+$3000 & 1.8 & HR~1318 \\
         & 2 & 160 & MN &        2$\times$2700 & 1.5 & HR~1318 \\
NGC~2855 & 1 & 120 & MJ &        2$\times$3600 & 2.0 & HR~2035 \\
         & 1 &  30 & MN &        3$\times$3600 & 3.0 & HR~2035 \\ 
NGC~3885 & 1 & 123 & MJ &        4$\times$3600 & 4.0 & HR~2429 \\
         & 1 &  33 & MN &        3$\times$3600 & 3.0 & HR~2429 \\
NGC~4224 & 1 &  57 & MJ &        2$\times$3600 & 2.0 & HR~2429 \\
         & 1 & 147 & MN &        2$\times$3600 & 2.0 & HR~2429 \\
NGC~4235 & 1 &  48 & MN &        2$\times$3600 & 2.0 & HR~2035 \\
         & 1 & 138 & MN &        2$\times$3600 & 2.0 & HR~2035 \\
NGC~4586 & 1 & 115 & MJ &        4$\times$3600 & 4.0 & HR~2701 \\
         & 1 &  25 & MN &        3$\times$3600 & 3.0 & HR~2701 \\
NGC~4984 & 1 &  90 & MJ & 3$\times$3600$+$3000 & 3.8 & HR~2035 \\
         & 1 &   0 & MN &        3$\times$3600 & 3.0 & HR~2035 \\
NGC 7049 & 2 &  57 & MJ &                 3600 & 1.0 & HR~2113 \\
         & 2 & 147 & MN &                 3600 & 1.0 & HR~2113 \\
NGC~7213 & 2 & 124 & MJ & 2$\times$3600$+$2700 & 2.8 & HR~2113 \\
         & 2 &  34 & MN &        2$\times$3600 & 2.0 & HR~2113 \\
NGC~7377 & 2 & 101 & MJ &        2$\times$3600 & 2.0 & HR~2113 \\
         & 2 &  21 & MN &        2$\times$3600 & 2.0 & HR~2113 \\
\noalign{\smallskip}   
\hline  
\noalign{\smallskip}   
\noalign{\smallskip}   
\noalign{\smallskip}   
\end{tabular}   
\begin{minipage}{18cm}  
NOTES. -- 
Col. 2: Observing run. 
Col. 3: Slit position angle measured North through East. 
Col. 4: Slit position. MJ = major axis; MN = minor axis. 
Col. 5: Number and exposure time of the single exposures. 
Col. 6: Total exposure time.
Col. 7: Giant star used as kinematical template in measuring 
        the stellar kinematics (see Sec. \ref{sec:datareduction}).
\end{minipage}  
\end{flushleft}  
\label{tab:log} 
\end{table*}

\subsection{Data reduction}
\label{sec:datareduction}

Basic data reduction was performed as in Corsini et al. (1999).  Using
standard {\tt ESO-MIDAS}\footnote{{\tt MIDAS} is developed and
maintained by the European Southern Observatory.}  routines, all the
spectra were bias subtracted, flat-field corrected by quartz lamp and
twilight exposures, cleaned of cosmic rays, and wavelength
calibrated. After calibration, the different spectra obtained for a
given galaxy along the same position angle were co-added using the
center of the stellar-continuum radial profile as a reference. The
contribution of the sky was determined from the outermost $\sim10''$
at the two edges of the resulting frames where the galaxy light was
negligible, and then subtracted.

The ionized-gas kinematics were measured from the \niipg, \ha\ and
\siipg\ emission lines by means of the {\tt MIDAS} package {\tt ALICE}
as done by Corsini et al.  (1999). The position and \FWHM\ of each
emission line were determined by interactively fitting one Gaussian to
each line plus a polynomial to its local continuum. The center
wavelength of the fitting Gaussian was converted into velocity in the
optical convention, then the standard heliocentric correction was
applied. The Gaussian \FWHM\ was corrected for the instrumental \FWHM ,
and then converted into the velocity dispersion. In the regions where
the intensity of the emission lines was low, we binned adjacent
spectral rows in order to improve the signal-to-noise ratio, $S/N$, of
the lines.
We expressed the velocity and velocity dispersion errors as a function
of the relevant line $S/N$ ratio as done in Corsini et al. (1999) and
derived ionized-gas velocity (\vg) and velocity dispersion (\sg)
as the weighted mean of values measured for the
different emission lines.
The kinematics of the ionized gas are reported in Table 3 and plotted
in Fig. \ref{fig:kinematics} for all the sample galaxies.

The stellar kinematics were obtained from the absorption features
present in the wavelength range running from about 4960 \AA\ to 5540
\AA\ and centered on the Mg line triplet
($\lambda\lambda\,5164,5173,5184$ \AA).  We used the Fourier
Correlation Quotient method (FCQ, Bender 1990) following the
prescriptions of Bender, Saglia \& Gerhard (1994).  The spectra were
rebinned along the spatial direction to obtain a nearly constant
signal-to-noise ratio larger than 20 per resolution element.  The
galaxy continuum was removed row-by-row by fitting a fourth to sixth
order polynomial as in Bender et al.  (1994). The stars adopted as
kinematical templates are given in Table \ref{tab:log}.  This allowed
us to derive, for each spectrum, the line-of-sight stellar velocity
(\vs) and velocity dispersion (\ss) by fitting a Gaussian to the
line-of-sight velocity distribution (LOSVD) at each radius. Velocities
were corrected for heliocentric velocity.
We derived errors on the stellar kinematics from photon statistics and
CCD read-out noise, calibrating them by Monte Carlo simulations as
done by Gerhard et al. (1998). These errors do not take into account
possible systematic effects due to template mismatch.
The kinematics of the stellar component are reported in Table 4 and
plotted in Fig. \ref{fig:kinematics} for all the sample galaxies.

\begin{figure*}[ht!]
\centering
\resizebox{\hsize}{!}{ 
\includegraphics[clip=true,angle=-90]{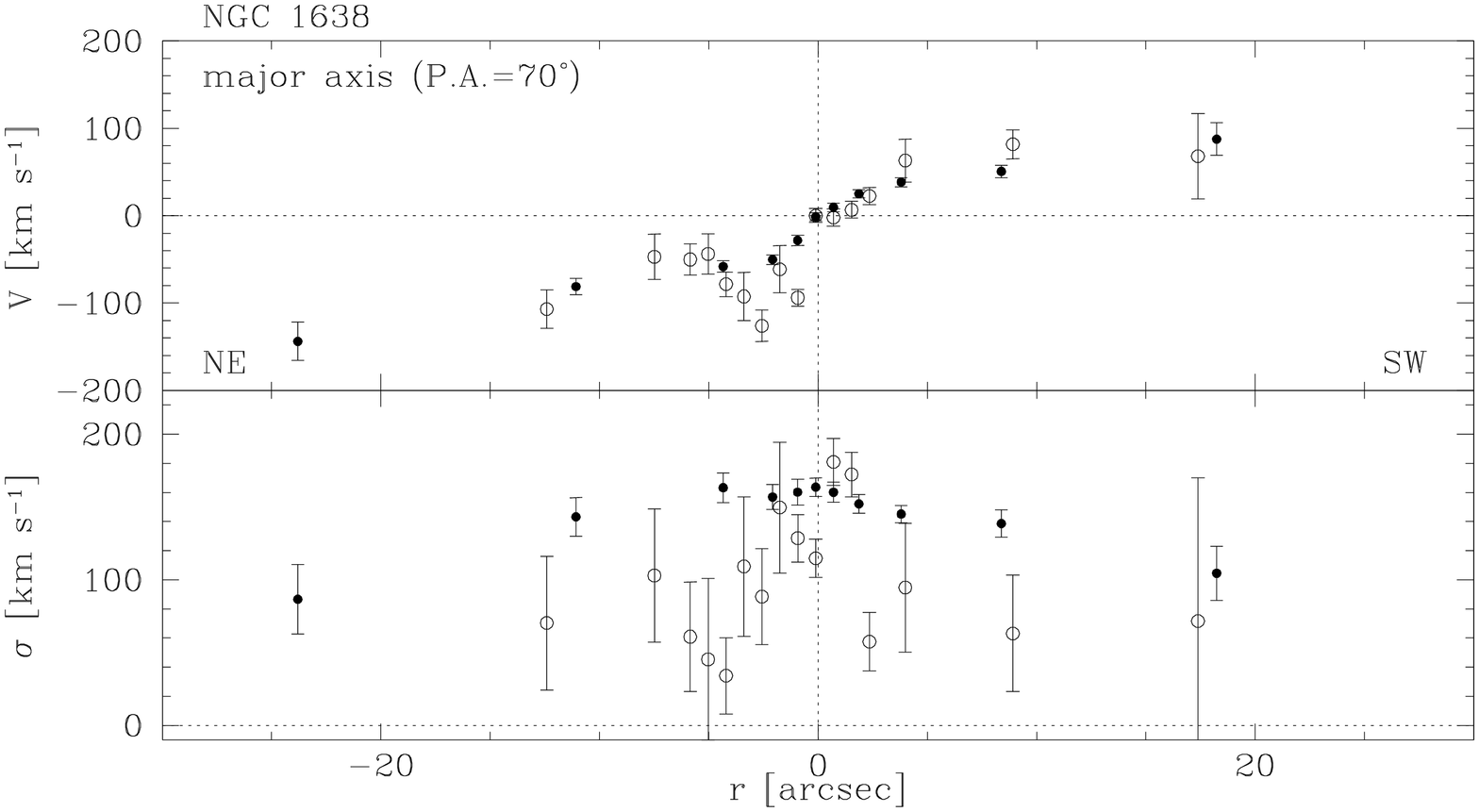} 
\includegraphics[clip=true,angle=-90]{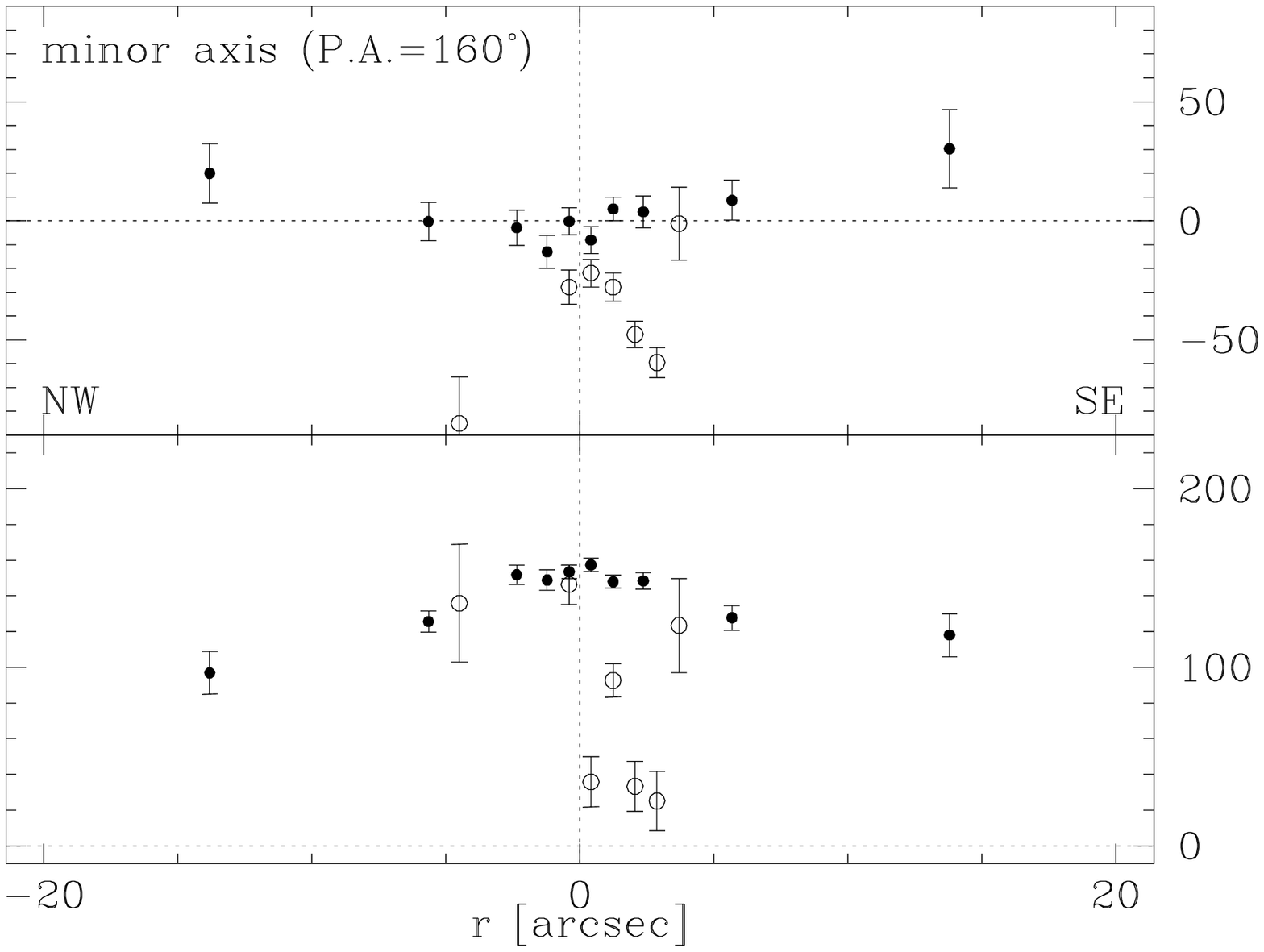}} 
\resizebox{\hsize}{!}{ 
\includegraphics[clip=true,angle=-90]{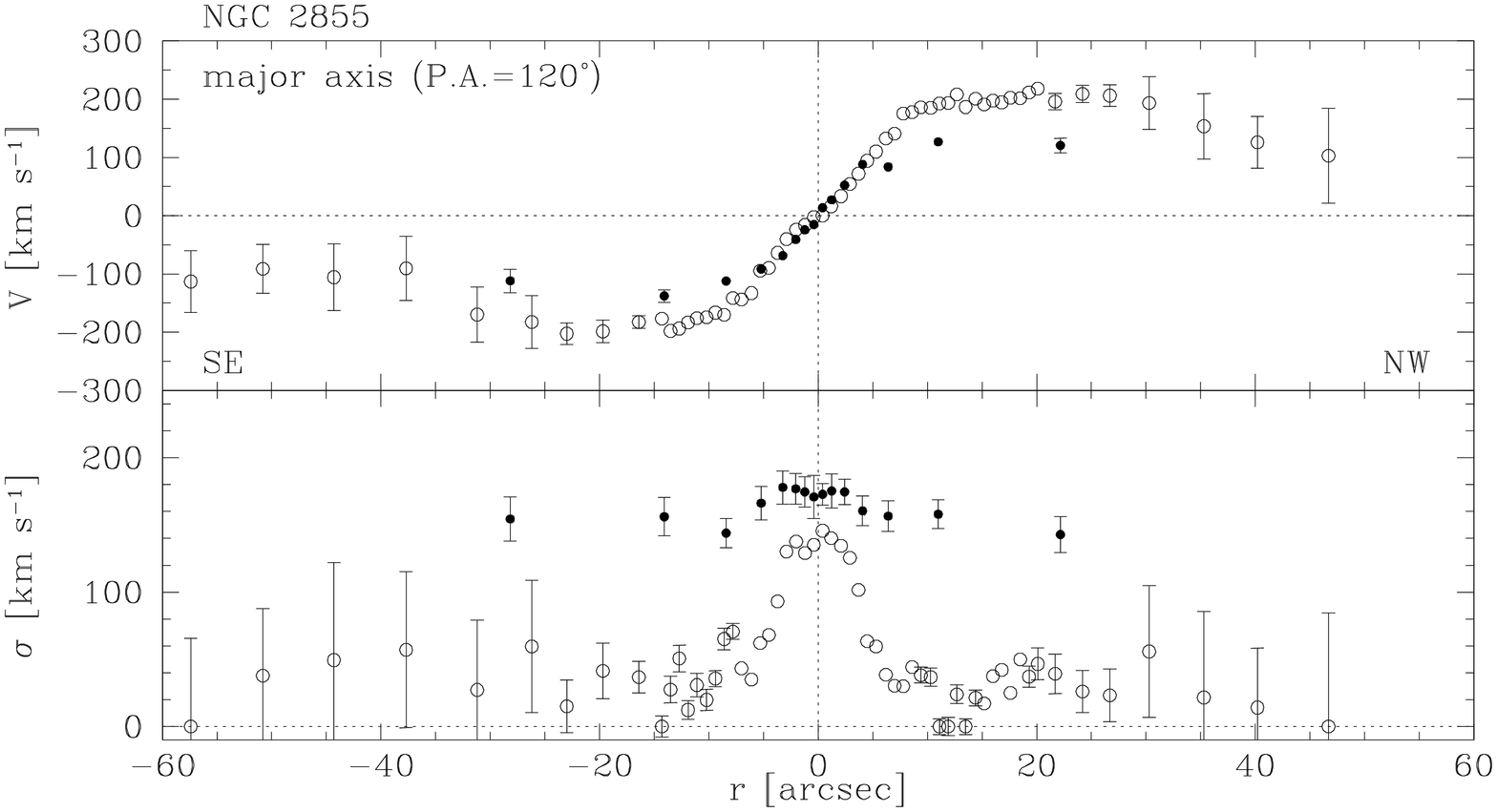} 
\includegraphics[clip=true,angle=-90]{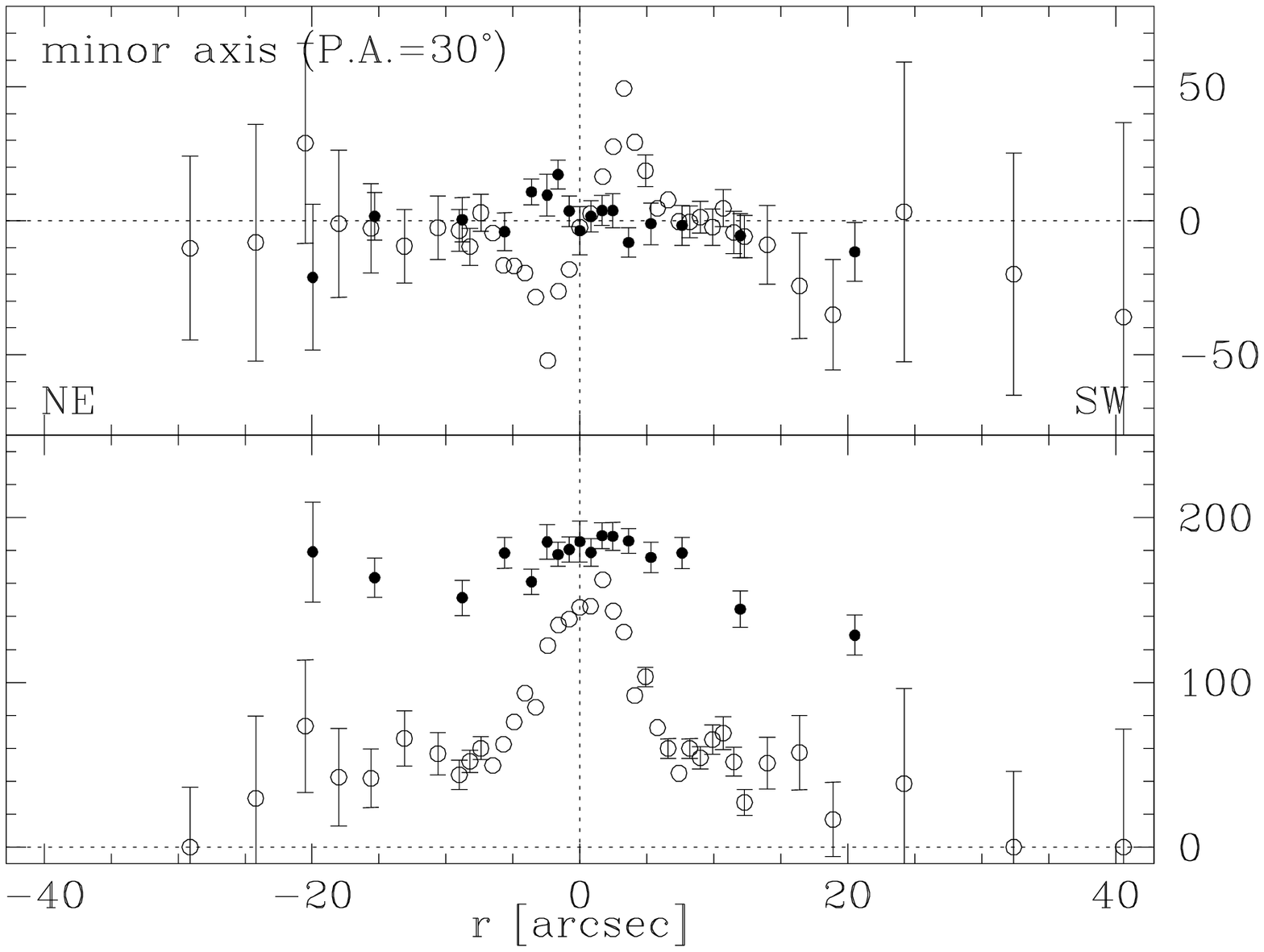}} 
\resizebox{\hsize}{!}{ 
\includegraphics[clip=true,angle=-90]{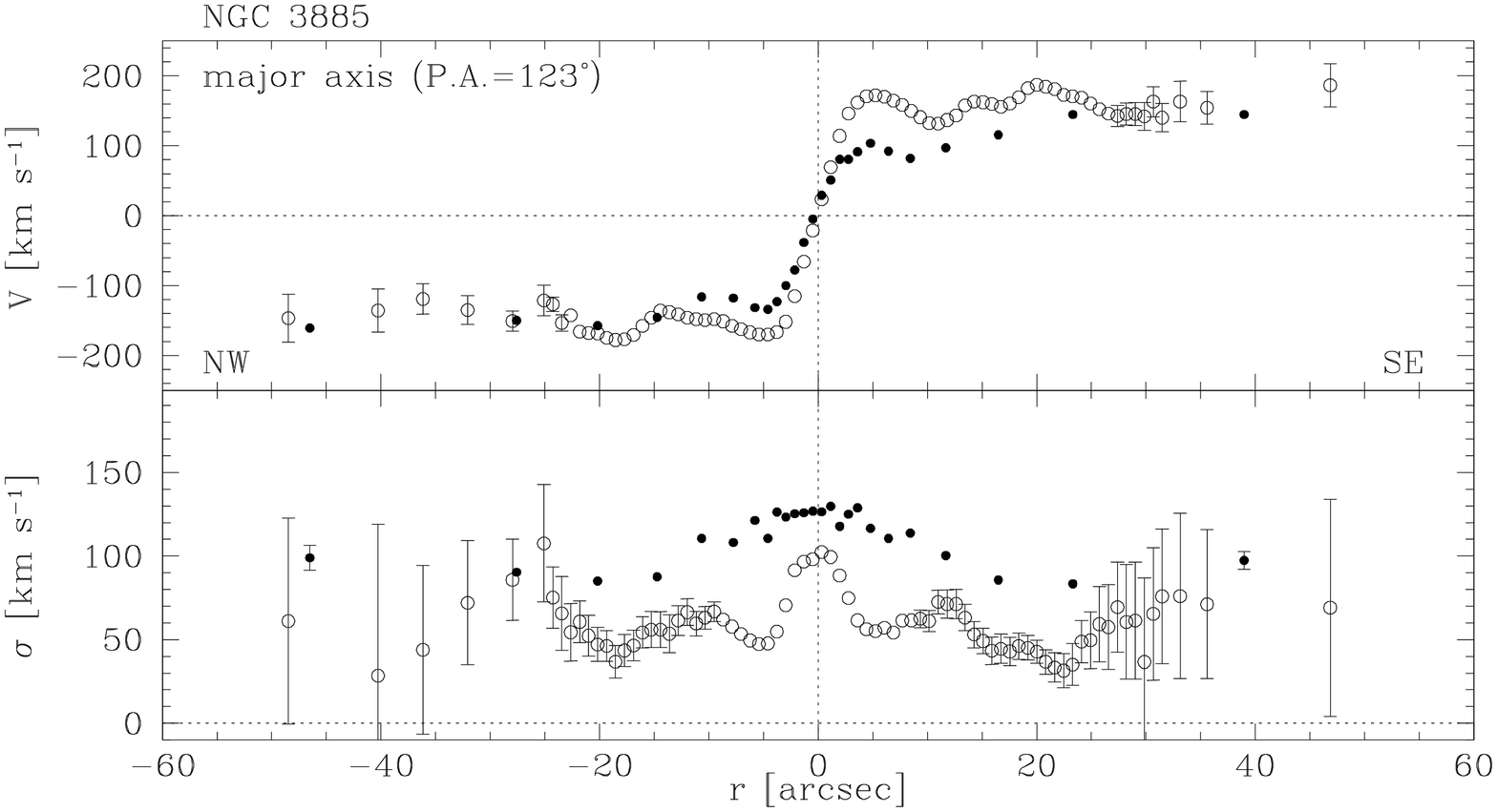} 
\includegraphics[clip=true,angle=-90]{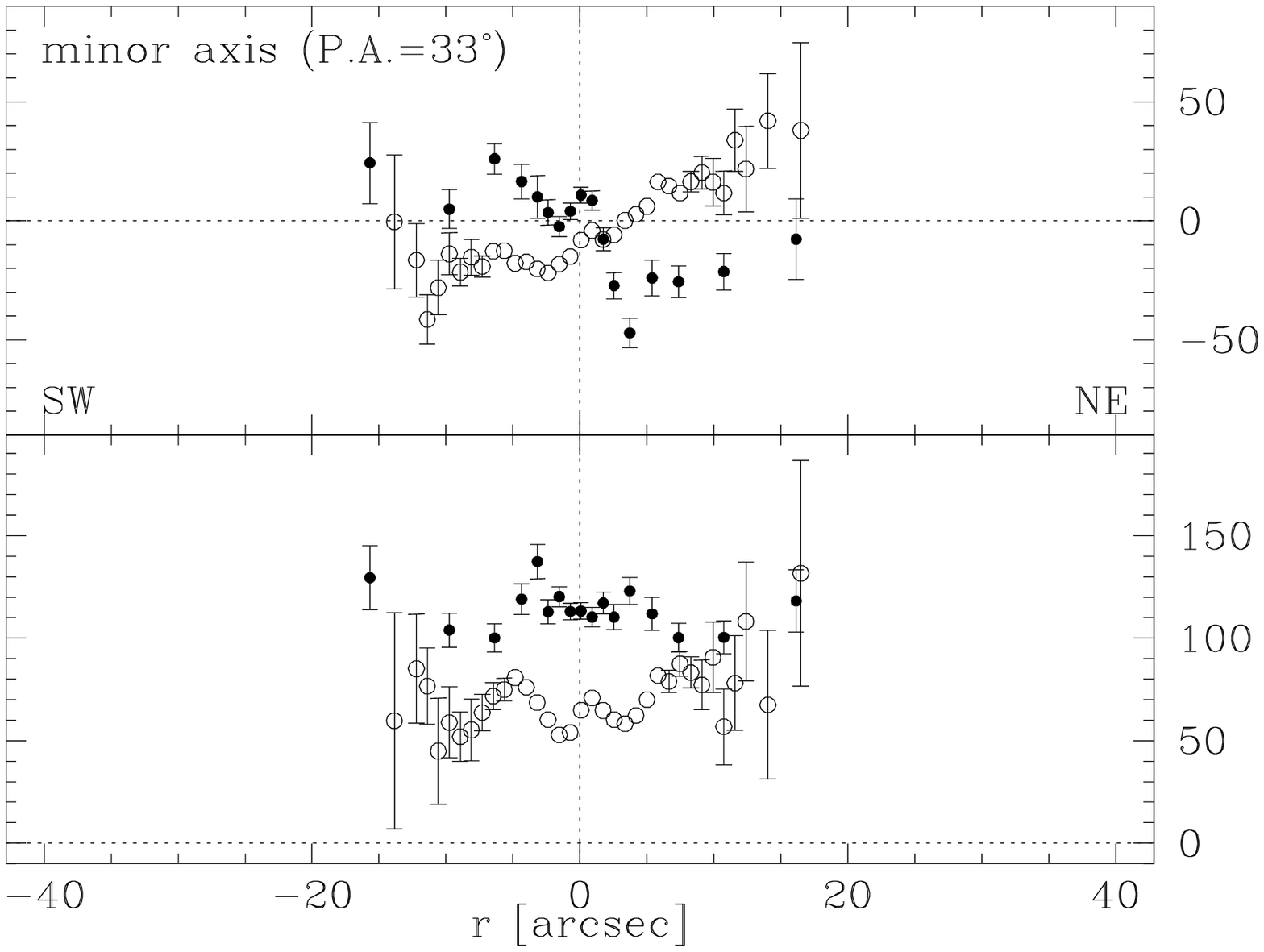}} 
\resizebox{\hsize}{!}{ 
\includegraphics[clip=true,angle=-90]{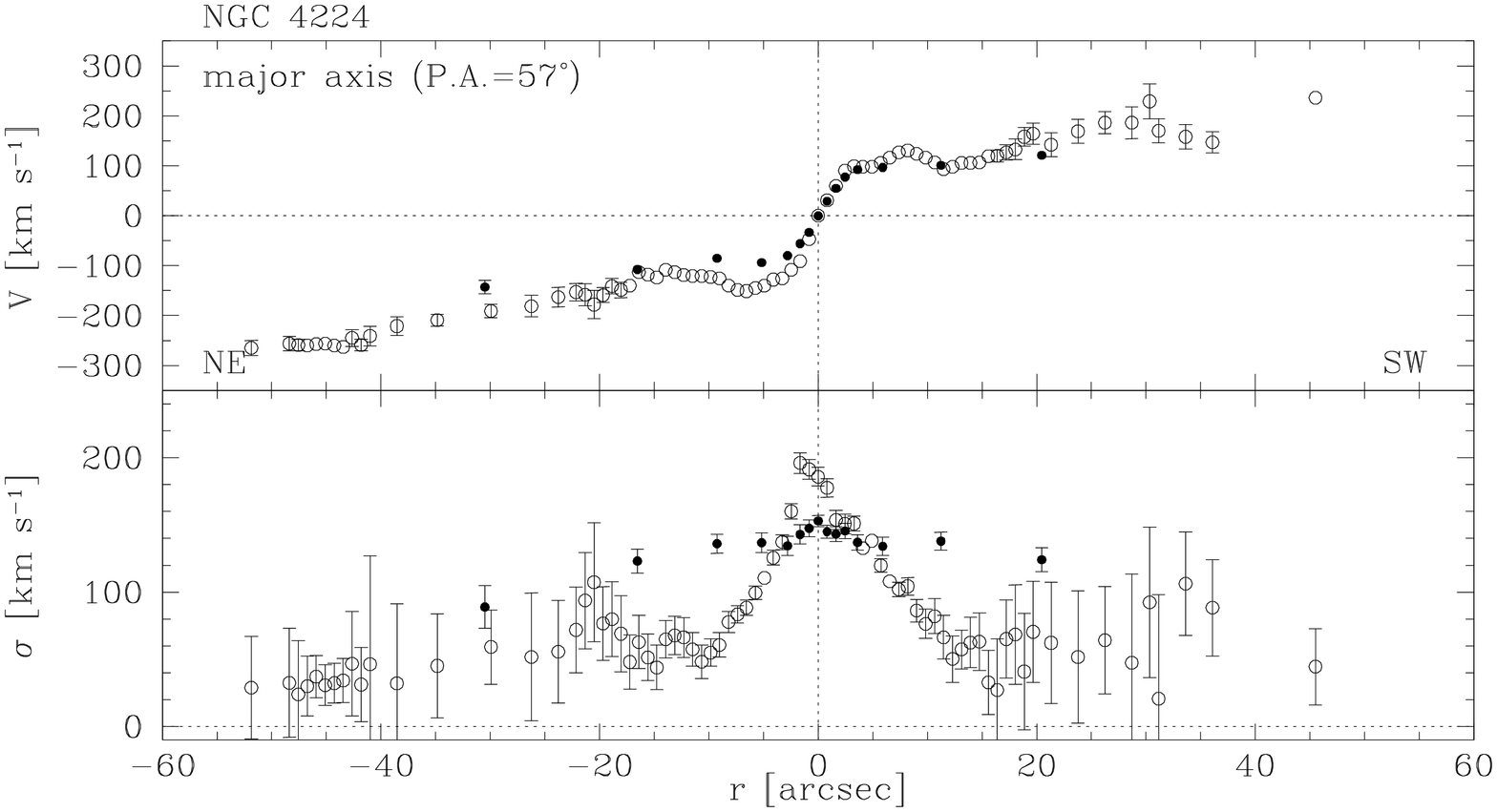} 
\includegraphics[clip=true,angle=-90]{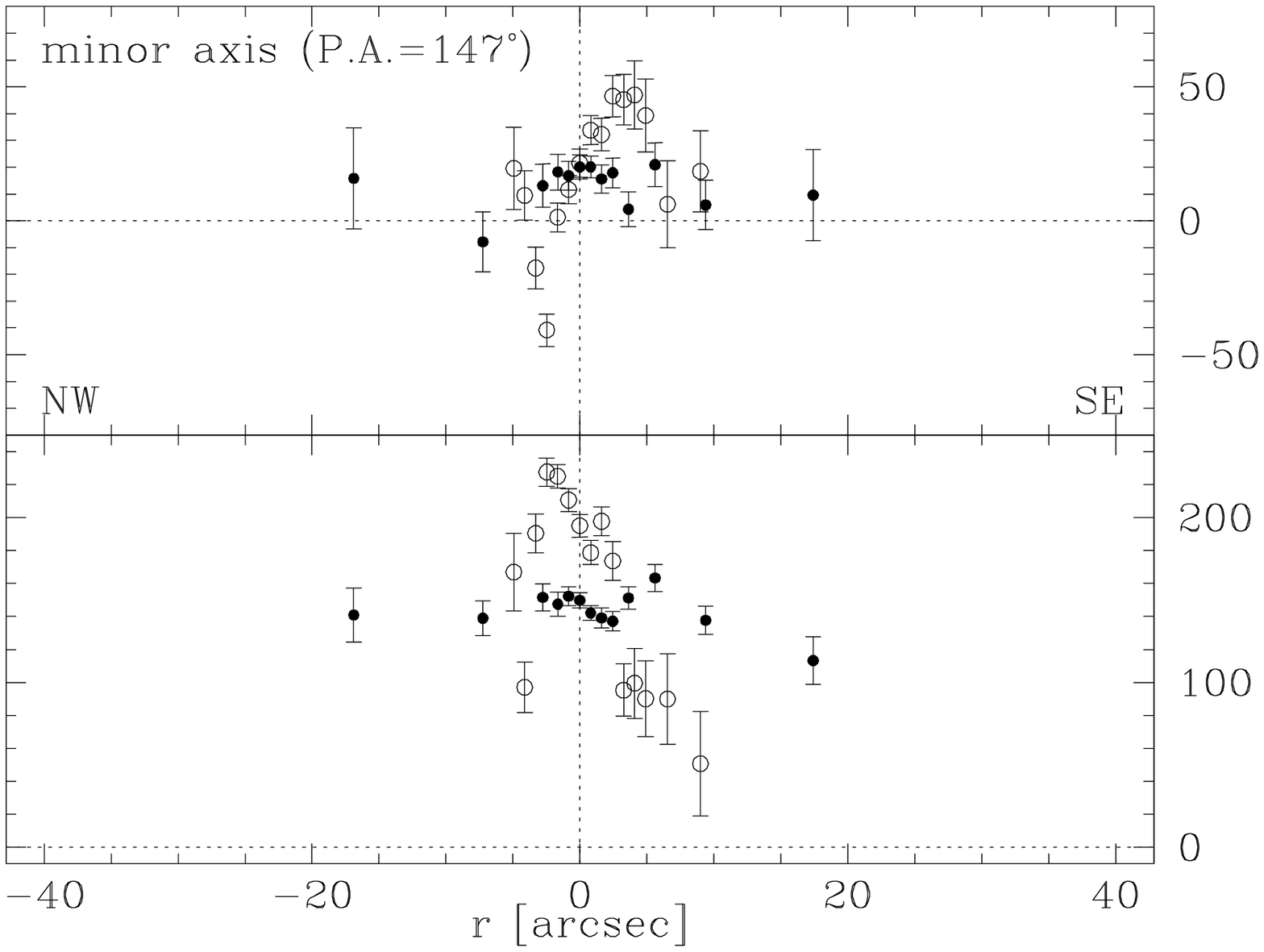}} 
\caption{Stellar ({\it filled circles\/}) 
  and ionized-gas ({\it open circles\/}) kinematics measured along
  major ({\it left panels\/}) and minor ({\it right panels\/}) axis of
  the sample galaxies. Errorbars smaller than symbols are not
  plotted.}
\label{fig:kinematics}
\end{figure*}

\addtocounter{figure}{-1}
\begin{figure*}[ht!]
\centering
\resizebox{\hsize}{!}{ 
\includegraphics[clip=true,angle=-90]{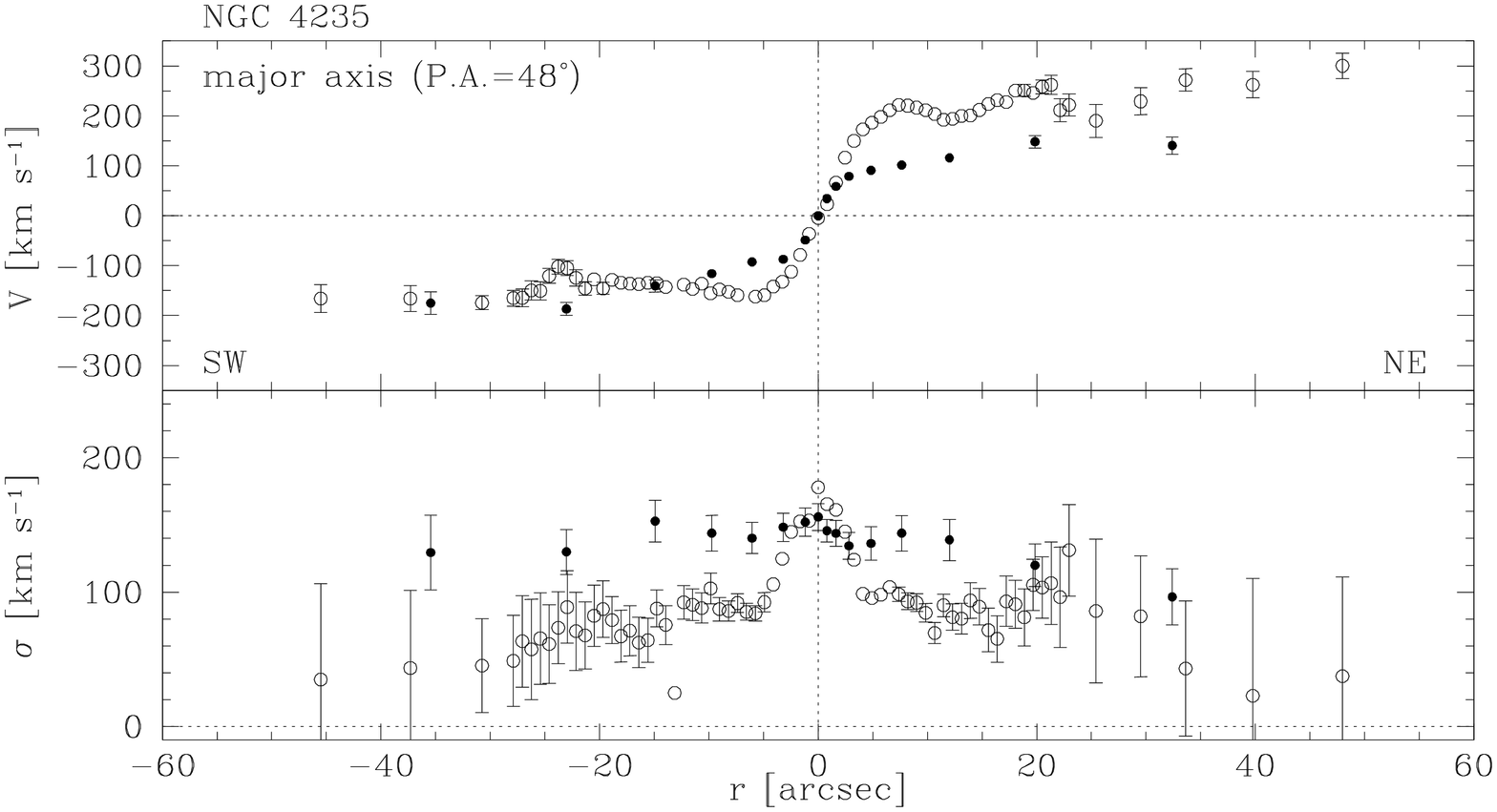} 
\includegraphics[clip=true,angle=-90]{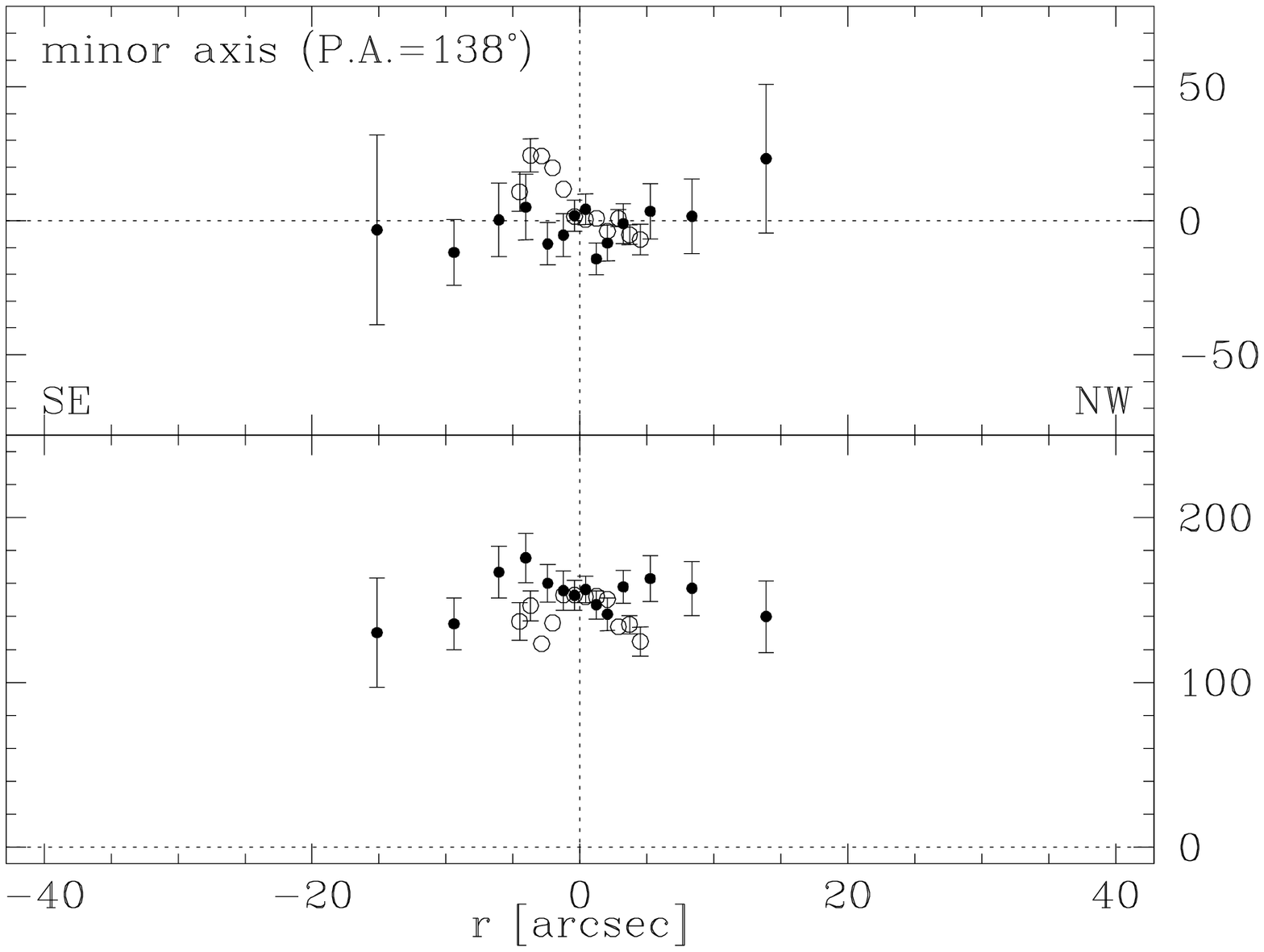}} 
\resizebox{\hsize}{!}{ 
\includegraphics[clip=true,angle=-90]{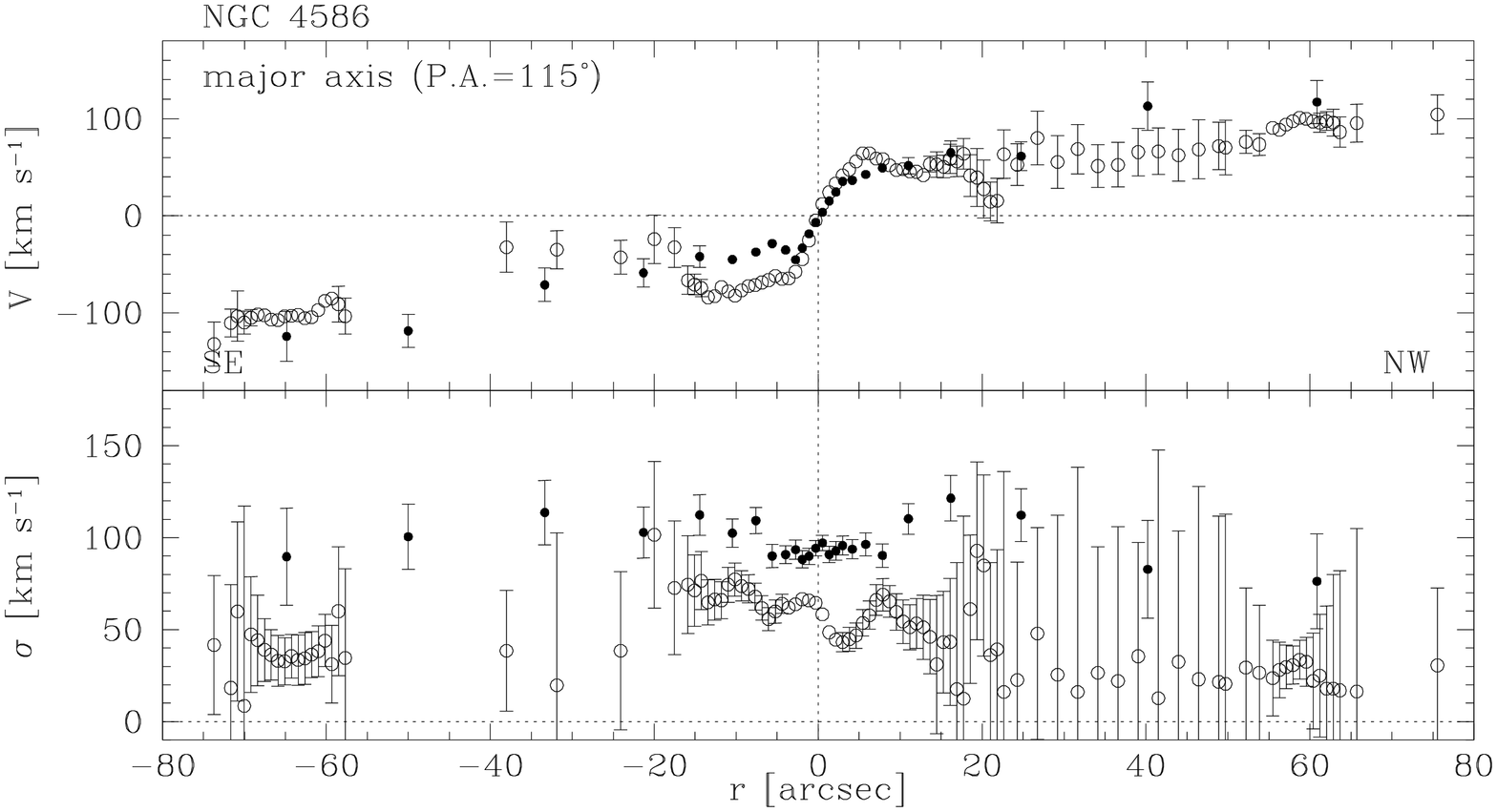} 
\includegraphics[clip=true,angle=-90]{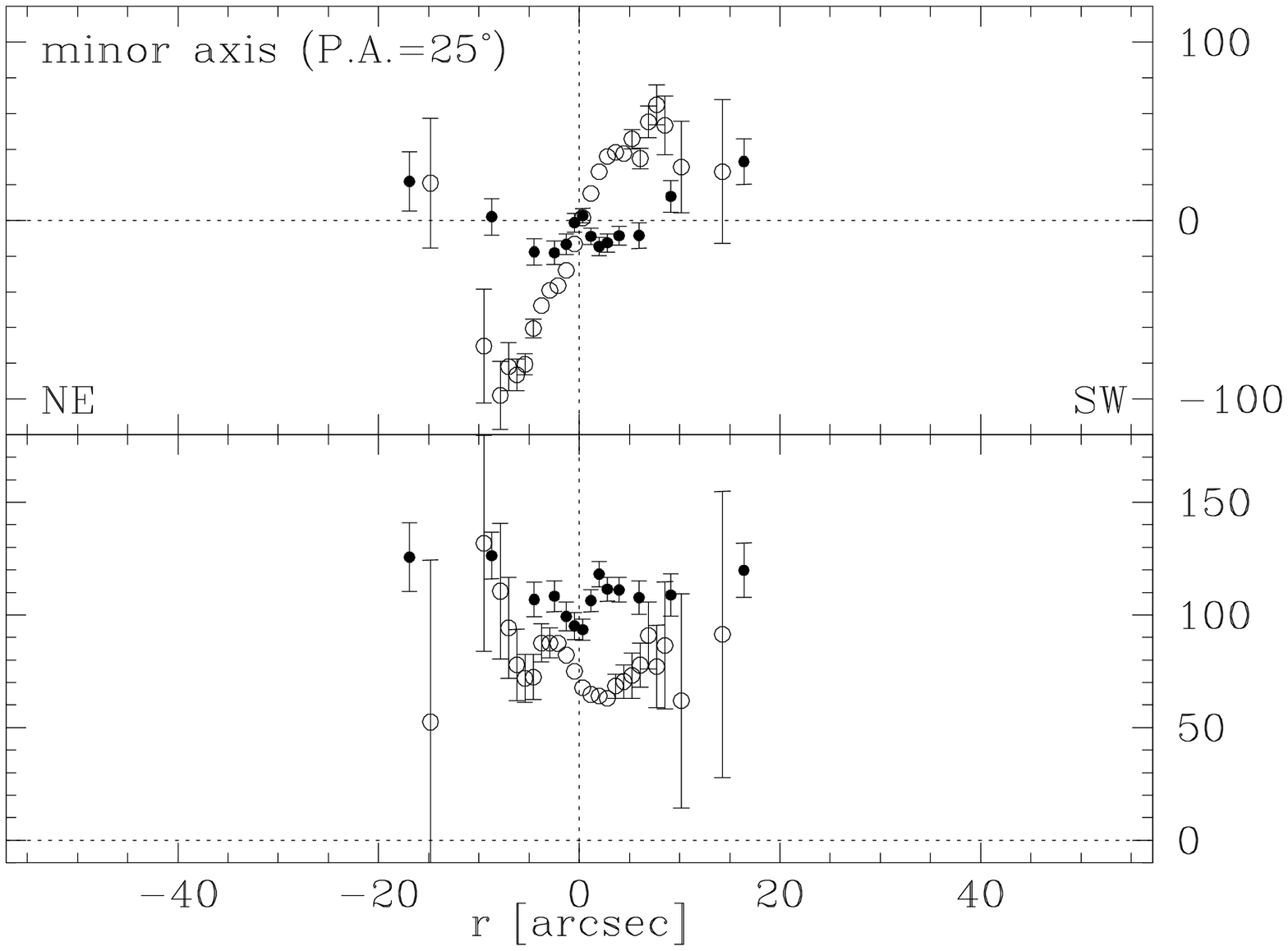}} 
\resizebox{\hsize}{!}{ 
\includegraphics[clip=true,angle=-90]{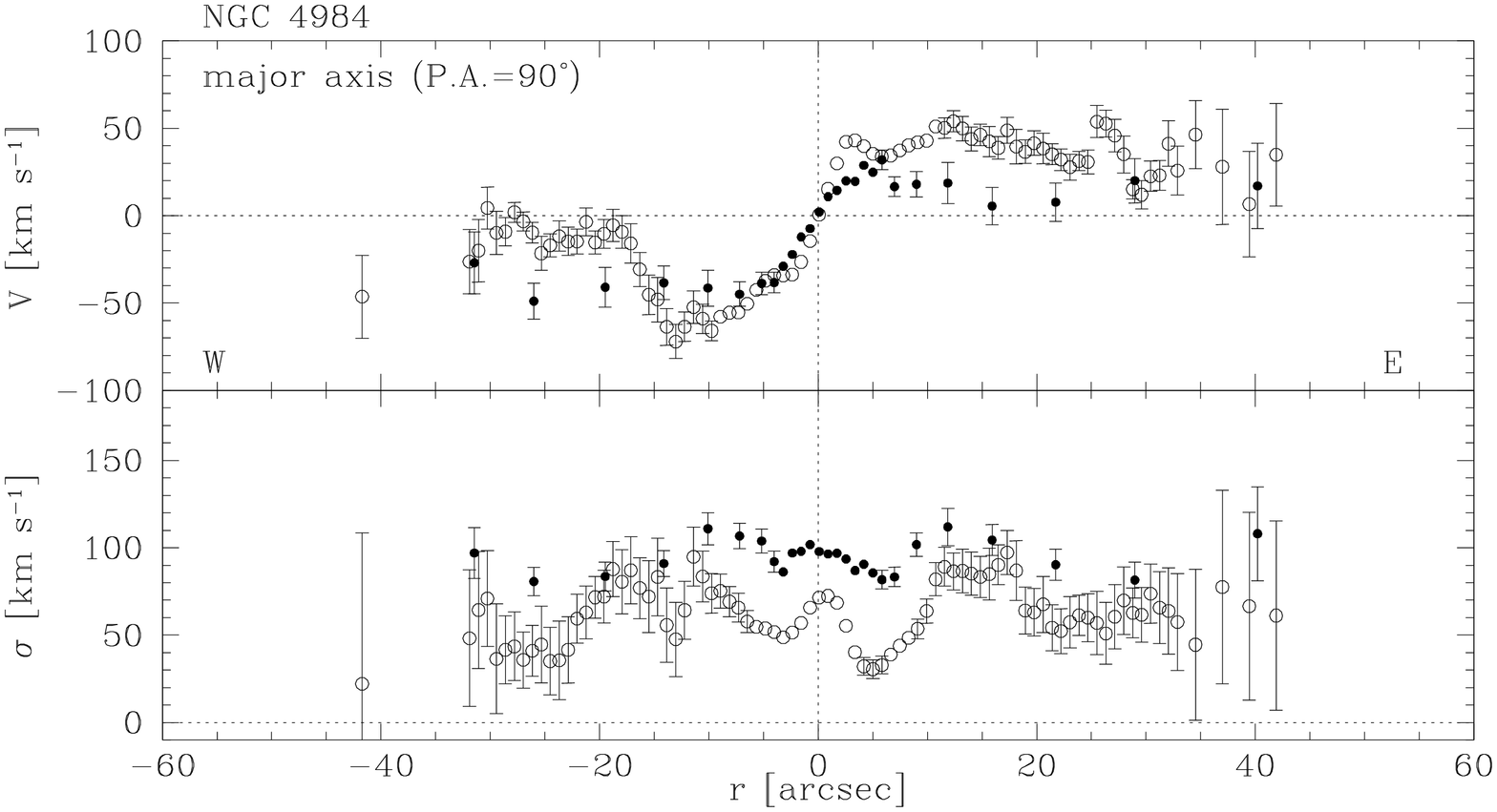} 
\includegraphics[clip=true,angle=-90]{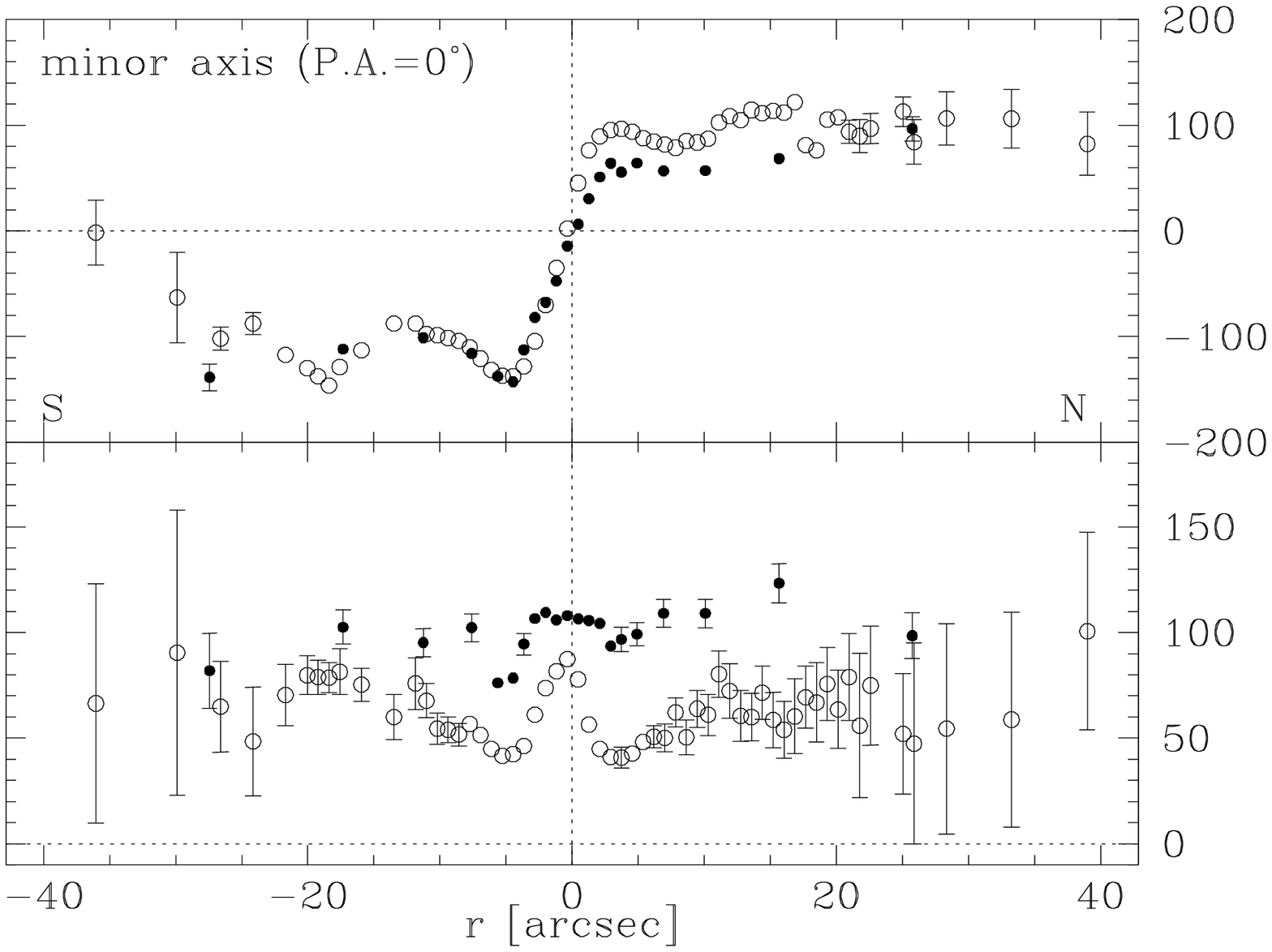}} 
\resizebox{\hsize}{!}{ 
\includegraphics[clip=true,angle=-90]{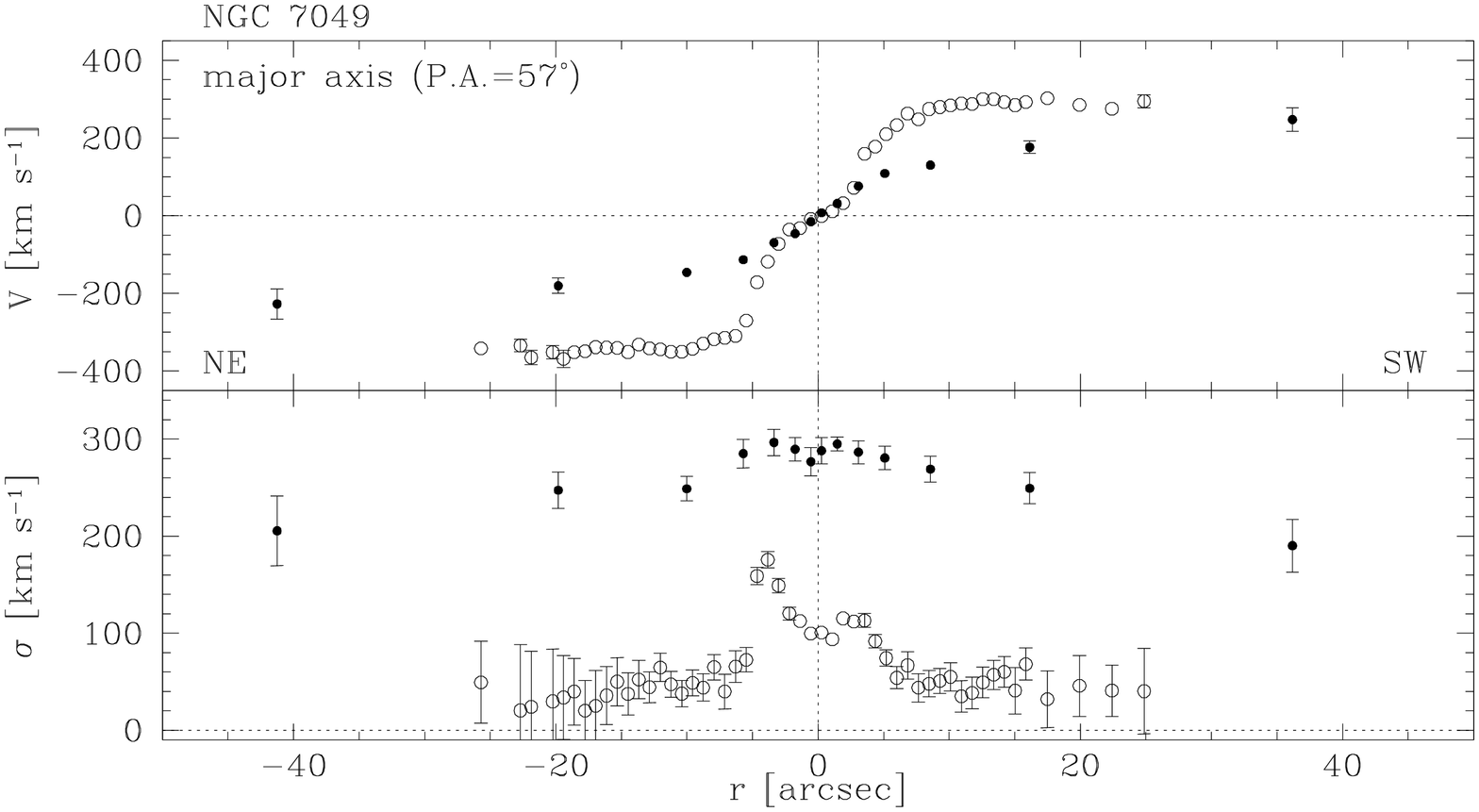} 
\includegraphics[clip=true,angle=-90]{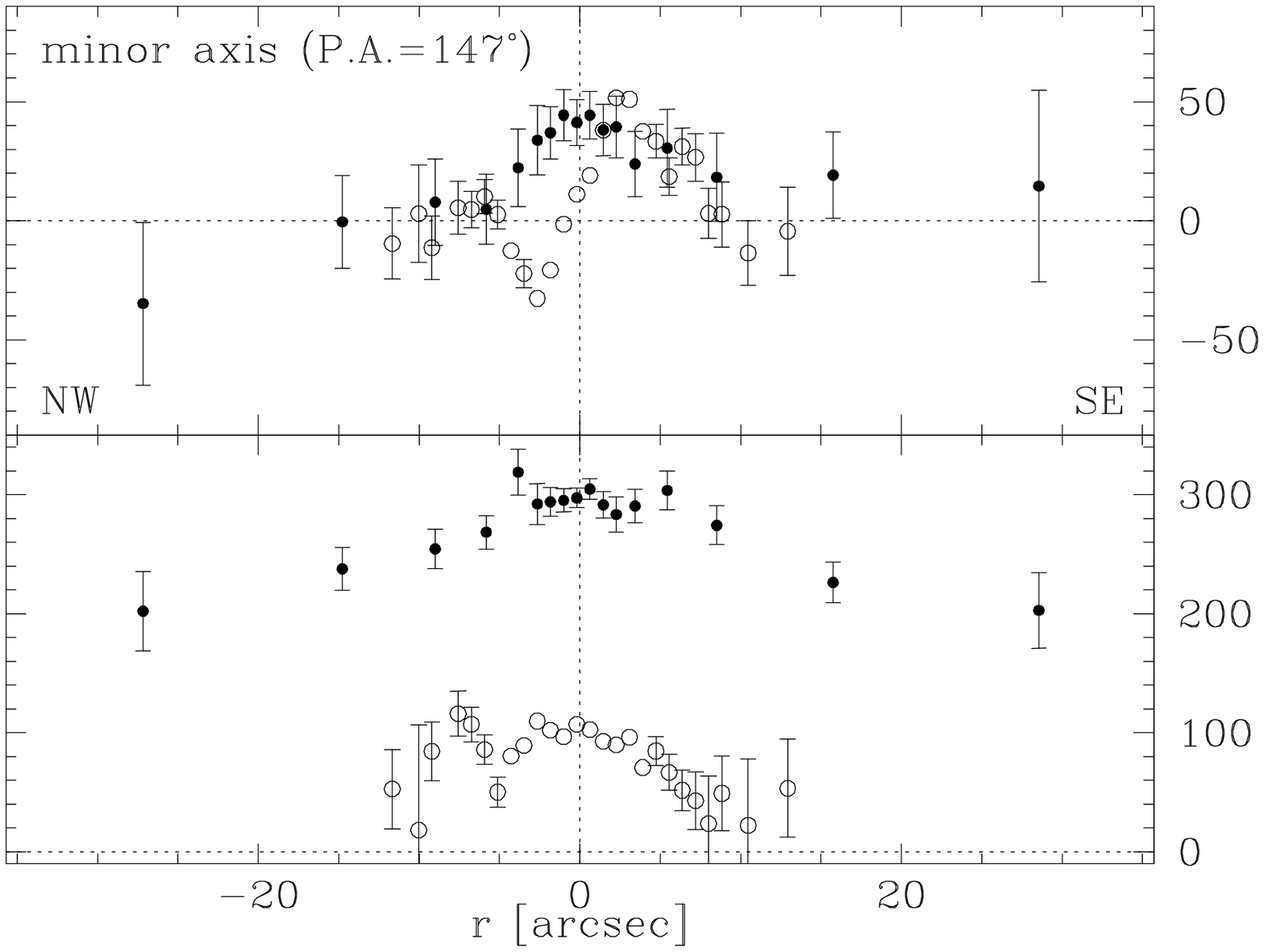}} 
\caption{(continued).} 
\end{figure*}

\addtocounter{figure}{-1}
\begin{figure*}[ht!]
\centering
\resizebox{\hsize}{!}{ 
\includegraphics[clip=true,angle=-90]{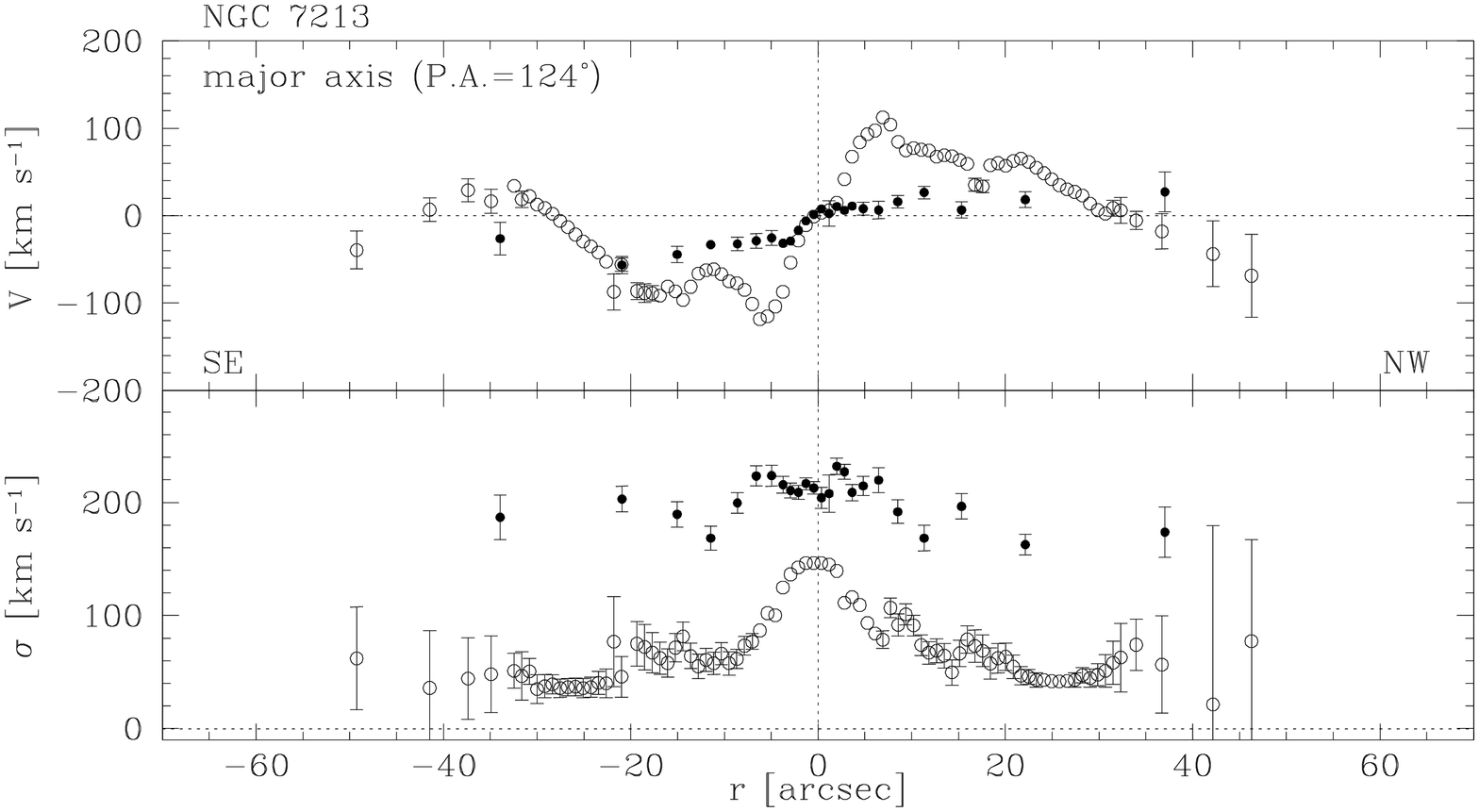} 
\includegraphics[clip=true,angle=-90]{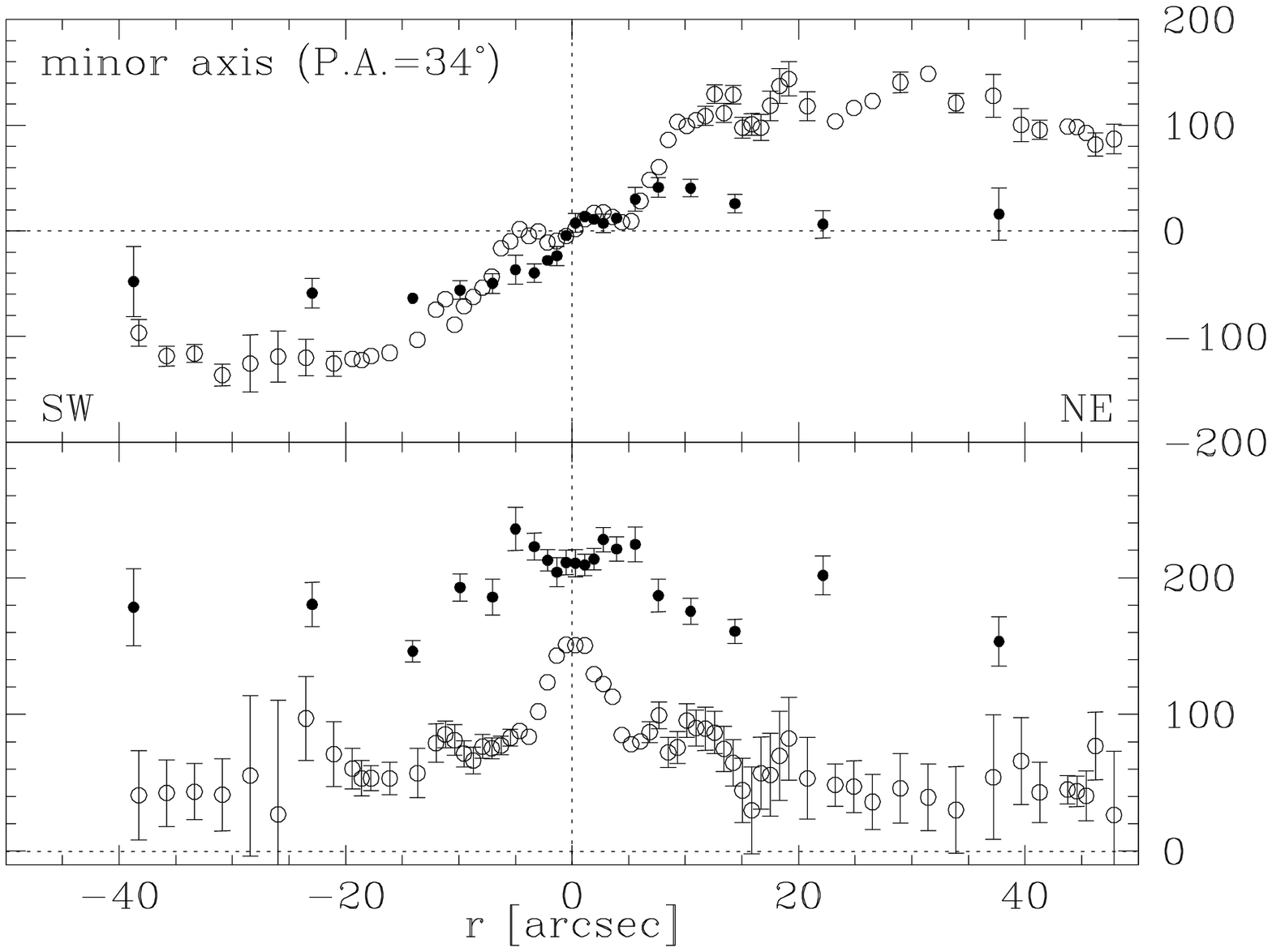}} 
\resizebox{\hsize}{!}{ 
\includegraphics[clip=true,angle=-90]{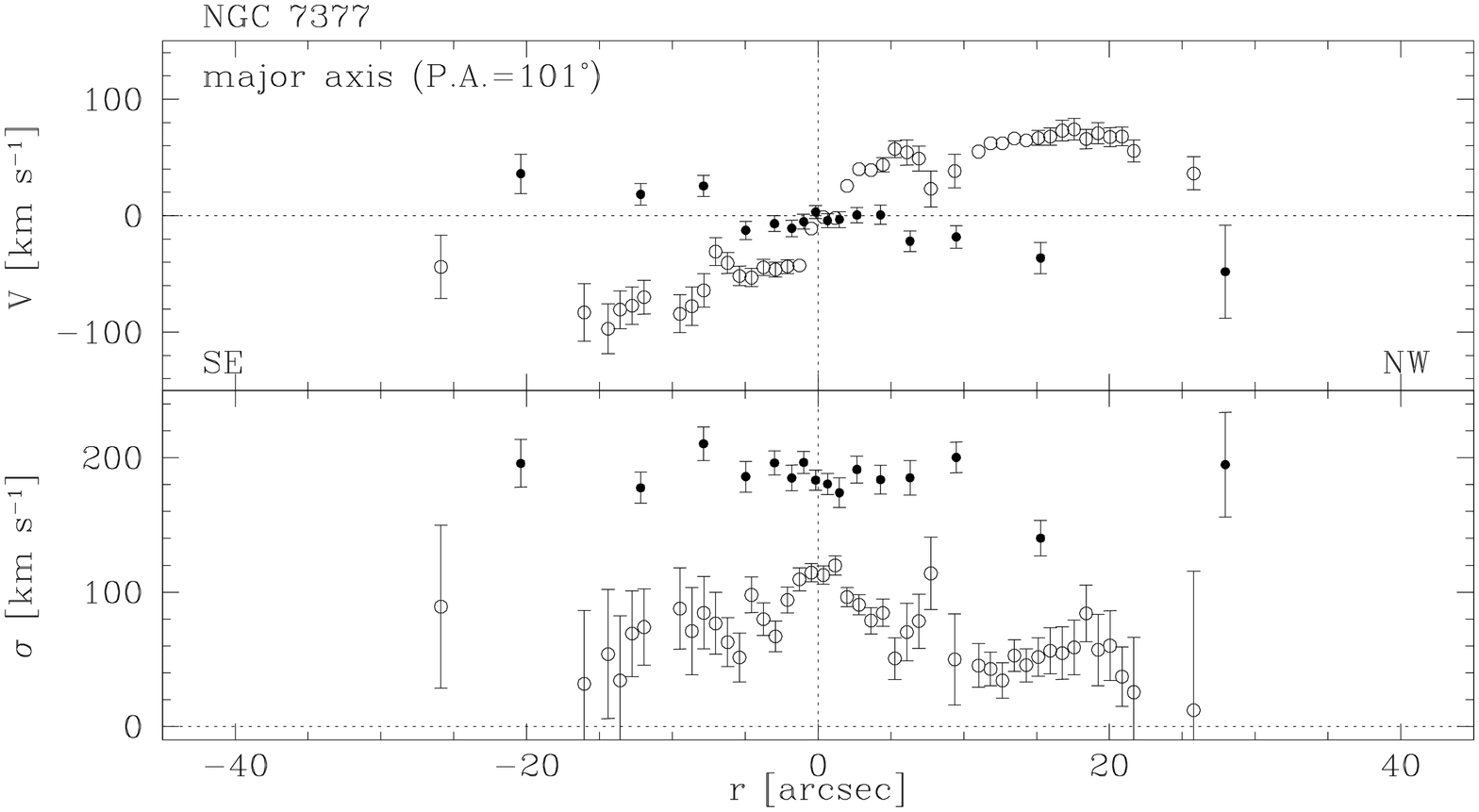} 
\includegraphics[clip=true,angle=-90]{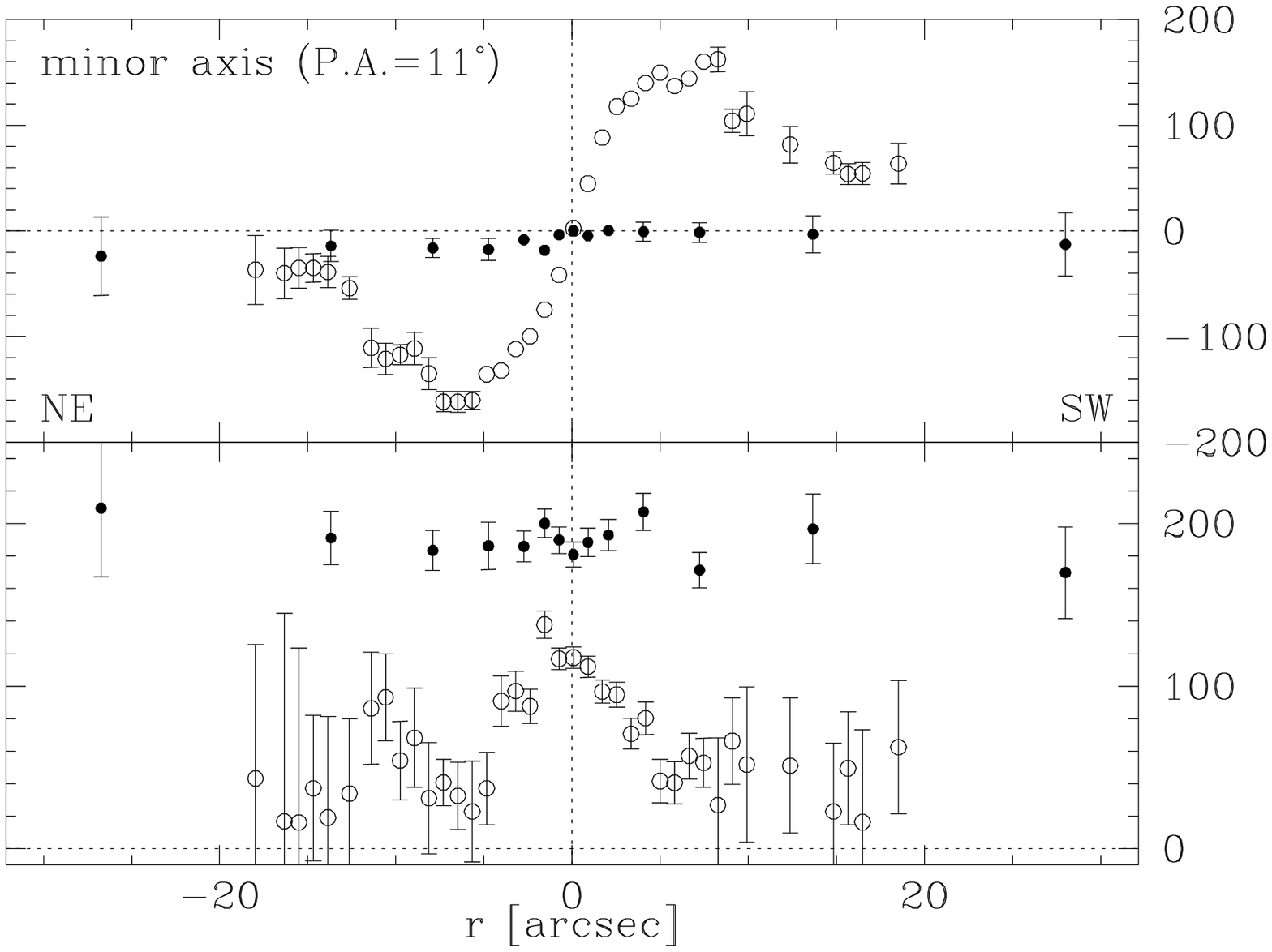}} 
\caption{(continued).} 
\end{figure*}

\section{The stellar and ionized gas kinematics} 
\label{sec:kinematics} 

The velocity curves and velocity dispersion profiles we measured for
the gaseous and stellar components along the major and minor axis of
the sample galaxies are presented in Fig. \ref{fig:kinematics}.
The rotation velocities of stars $V_\star$ ($\equiv
|v_\star-V_\odot|$) and ionized gas $V_g$ ($\equiv |v_g-V_\odot|$)
given in Table 3 and 4 and plotted in Fig.  \ref{fig:kinematics} are
the observed velocities after subtracting the systemic velocities of
Table 1 and without applying any correction for galaxy inclination.
$\sigma_\star$ and $\sigma_g$ are the velocity dispersion of stars and
gas, respectively.
The case of NGC 2855 is shown in Fig. \ref{fig:kinematics} for the
sake of completeness. For details the reader is referred to Corsini et
al. (2002).

\subsection{Major-axis kinematics}

The only noteworthy features observed in the major-axis kinematics of
the ionized-gas and stellar component are the counterrotation of gas
with respect to stars found in NGC 7213, and the reversal of gas
rotation measured in the outermost regions of NGC 7377. In the sample
galaxies \Vs\ has always a shallower gradient than \Vg\ (except for
the inner regions of NGC 2855 and NGC 7049, where \Vs$>$\Vg ), and
\Vs$\ap$\Vg\ at the last observed radius (except for NGC 7213 and NGC
7377). Typically \ss\ exceeds 150 \kms\ and it is larger than \sg\ at
all radii (except in the center of NGC 4224 and NGC 4235 where
\ss$\la$\sg ). If we took into account only the major-axis data, we
could straightforwardly explain the observed kinematical properties by
a dynamical model where gas is confined in the disk and supported by
rotation and stars mostly belong to the bulge and are supported by
dynamical pressure.

\subsection{Minor-axis kinematics}

This is not the case if we consider the minor-axis kinematics too.
The minor-axis velocity gradient we measure for the gaseous component
of 8 out of 10 sample galaxies is hardly explained if we naively
assume that the ionized gas is moving onto circular orbits in the
galaxy disk.

Non-zero gas velocities are measured in the nuclear regions along the
minor axis of NGC 2855, NGC 3885, NGC 4224, NGC 4586, and NGC 7049.
The gas velocity curve shows a steep gradient rising to a maximum
observed velocity of $\ap50$ \kms\ in the inner few arcsec, then
the velocity drops to $\ap0$ \kms\ at larger radii.
Non-zero gas velocities are measured along the minor axis of NGC 4984,
NGC 7213, and NGC 7377 all over the observed radial range. The maximum
velocity is $\ga150$ \kms\ and it is even larger than the maximum
value measured along the major axis.
Finally, along the minor axis of the remaining 2 sample galaxies,
namely NGC 1638 and NGC 4235, the gas is either too poorly detected or
it shows a velocity curve that is too asymmetric to derive any
conclusion.  We will not consider these 2 galaxies further.

As far as the minor-axis stellar kinematics is concerned, no
significant rotation velocity is observed along the minor axes of NGC
1638, NGC 2885, NGC 4235, and NGC 7377.
The same is true for NGC 4224 and NGC 7049 if we interpret the falling
of the stellar velocity from a central maximum of $\ap30$ \kms\ to $0$
\kms\ off the center as due to an offset position of the slit of
$\ap1''$ with respect to the nucleus and parallel to the minor axis.
Non-zero stellar velocities are measured along the minor axis of NGC
4984 and NGC 7213 over all the observed radial range. For these
galaxies the amplitude of the stellar velocity curve measured along
the minor axis is larger than that measured along the major axis.
Stars are counterrotating with respect to gas along the minor axis of
NGC 3885.

\section{Discussion and conclusions} 
\label{sec:discussion}

According to the ionized-gas velocity field, we distinguish three
classes of objects in sample galaxies:

\begin{enumerate}

\item In NGC 4984, NGC 7213, and NGC 7377 an overall gas velocity
curve is observed along the minor axis without zero-velocity points,
out to the last measured radius. These gas kinematics have been
interpreted as due to the warped inner structure of the gaseous disk.

\item In NGC 3885, NGC 4224, and NGC 4586 the gas velocity rises
almost linearly in the inner regions of both the major (where gas
rotates faster than stars) and minor axes. Non-zero gas velocities
along the minor axis are confined in the nuclear regions. Such gas
kinematics have been explained as being due to non-circular motion
induced by a triaxial potential.
  
\item In NGC 2855 and NGC 7049 the major-axis gas velocity gradient in
the innermost $\ap500$ pc (where gas rotates slower than stars) is
shallower than that measured farther out (where gas rotates faster
than stars). Moreover the non-zero gas velocities observed along the
minor axis are confined in the nuclear regions. These kinematic
features have been attributed to the presence of an inner polar disk
of gas.
 
\end{enumerate}

Each of the above categories is discussed in detail in the following
sections.

\subsection{Warped gaseous disks}
\label{sec:warp}

An inner warp of the gaseous disk can account for the gas velocity
curve observed along the minor axes in the low-inclined galaxies NGC
4984, NGC 7213, and NGC 7377.

NGC 4984 is classified as a very early-type Sa spiral in CAG and as a
weakly-barred S0/a in RC3. Sandage \& Bedke (1994) pointed out that
this galaxy has a peculiar structure characterized by two sets of
spiral arms. The inner spiral pattern is made by multiple dust arms
(see the inset of Panel 73 in CAG) and shows a strong variation of the
isophotal position angle from $\ap20^\circ$ to $\ap90^\circ$ in the
innermost $30''$. This feature results from the near-infrared
photometry by Jungwiert, Combes \& Axon (1997) who interpreted it as
the possible signature of a double bar. A second set of spiral arms is
visible outside the main body of the galaxy both in Panel 73 of the
CAG and in the DSS plate reprinted in Fig.  \ref{fig:sample}.  The
position angle of the outermost arms is $\ap25^\circ$. This change of
position angle is associated with a variation of the inclination from
$i\ap73^\circ$ at $\ap80''$ to $i\ap56^\circ$ at $\ap130''$, and it is
indicative of a warped disk. Therefore NGC 4984 has a peculiar
structure with a double bar in the central $30''$ and a warped disk at
larger radii.  To perform our spectroscopic observations we adopted
for the position angle of the major axis $\rm P.A. = 90^\circ$. This
value is given by RC3. Our kinematic data extend out to $\ap40''$ from
the galaxy center on both the observed axes ($\rm P.A.  = 0^\circ,
90^\circ$), which do not correspond to the actual photometric major
and minor axis of the galaxy in the measured radial range.

A geometrical decoupling between the gaseous and stellar disks
characterizes NGC 7213 in the radial region we covered with our
spectroscopic observations ($\la40''$).  Indeed, the orientation of
the circumnuclear ring-like structure observed in the map of the
H$\alpha$ emission obtained by Hameed \& Devereux (1999) does not
coincide with that of the stellar disk as derived from the
near-infrared photometry by Mulchaey, Regan \& Kundu (1997). The
ionized-gas ring has a diameter of $\ap21''$, a major-axis position
angle $\rm P.A.\ap40^\circ$ and an inclination
$i\ap30^\circ$. According to Mulchaey et al. (1997) the isophotal
ellipticity is about zero out to $\ap40''$, suggesting that in this
region the stellar disk is almost face-on with a major-axis position
angle of $\ap80^\circ$.  This is consistent with our kinematic data,
since we measure small rotation velocities for the stellar component
(\Vs$\la50$ \kms ) along both the observed axes ($\rm P.A.  =
34^\circ, 124^\circ$). These axes can be actually considered as
intermediate axes. At larger radii the inclination and the major-axis
position angle of the disk are $i=27^\circ$ and $\rm P.A. =
124^\circ$, respectively (see Table 1).
A giant filament of ionized gas is located at $\ap20$ kpc south of the
galaxy (Hameed \& Devereux 1999) and the HI velocity field reveals NGC
7213 to be a highly disturbed system undergoing an acquisition event
(Hameed et al. 2001).  Therefore we conclude that the kinematic and
photometric properties of NGC 7213 are consistent with a warped
gaseous disk in the context of an ongoing merge process.

According to Sandage \& Bedke (1994) NGC 7377 is one of the most
unusual galaxies in their atlas. They state that there is no evidence
of recent star formation over an otherwise normal S0 disk, yet the
entire disk is threaded by a multi-armed spiral pattern composed of
dust lanes only. These lanes are more visible to the south-west (Panel
75, CAG).
The morphology and the kinematics of NGC 7377 (Fig.
\ref{fig:kinematics}) are similar to those of NGC 7213.  In particular
the major and minor-axis gas kinematics of NGC 7377 behave very
similarly to the minor and major-axis gas kinematics of NGC 7213,
respectively. So again a warp of the gaseous disk can be suspected.
We interpret the counterrotation of gas and stars observed along the
galaxy major axis and the gas velocity decrease along the minor axis
(which is steeper than Keplerian) as due to a geometrical decoupling
between the gaseous and stellar disks, which are not coplanar.

\subsection{Non-circular gas motions in a triaxial bulge or in a bar}
\label{sec:triaxiality}

Non-zero gas velocities are measured along the minor axis in the
bulge-dominated region of NGC 3885, NGC 4224, and NGC 4586, and a
velocity gradient is observed along their major axis too. We interpret
the observed kinematics as due to non-circular gas motion in the
principal plane of the disk caused by bulge triaxiality or due to the
action of a bar.
Indeed the intrinsic shape of bulges is generally triaxial and the
gaseous disk lies on the principal plane perpendicular to the bulge
short axis (Bertola et al. 1991). Gas settled onto a principal plane
of a non-rotating triaxial potential moves onto closed elliptical
orbits. These orbits become nearly circular as soon as the distance
from the center increases (de Zeeuw \& Franx 1989; Gerhard et
al. 1989).  The same is true in barred potentials too (Athanassoula
1992).
The central velocity gradients measured along both major and minor
axis of NGC 3885, NGC 4224, and NGC 4586 are due to the orientation of
the inner elliptical orbits lying on the disk plane and seen at
intermediate angle between their intrinsic major and minor
axes. Minor-axis gas rotation velocity drops to zero where elliptical
orbits become circular.

The triaxiality of the bulge of NGC 3885 is suggested by the
significant Lindblad's misalignment (Lindblad 1956) between the
apparent major axes of bulge and disk ($\rm \Delta P.A.\ap10^\circ$).
The bulge-to-disk misalignment measured on the photographic plates of
NGC 3885 available in CAG (Panel 76 and 87) is consistent with the
isophotal twist found in the near infrared by Jungwiert et al. (1997)
in spite of the presence of the dust lanes threading through the
smooth disk and crossing the bulge close to its center. Therefore the
twist between the isophotes of bulge and disk of NGC 3885 cannot be
attributed to the bulge obscuration by dust.

Both NGC 4224 and NGC 4586 have a boxy/peanut-shape bulge (Panel 76,
CAG). These bulges can be interpreted as bars seen nearly edge on
(Kuijken \& Merrifield 1995; Bureau \& Athanassoula 1999).  This
supports the idea that the gas velocity gradient observed along their
minor axis is due to the non-circular gas motion in a triaxial
potential. This is in agreement with the recent results by
Athanassoula \& Bureau (1999), who studied by means of hydrodynamical
simulations the appearance of the position-velocity diagram of the
gaseous component in edge-on barred galaxies as a function of the
viewing angle.

\begin{figure*}[ht!]
\centering
\resizebox{\hsize}{!}{ 
\includegraphics[clip=true]{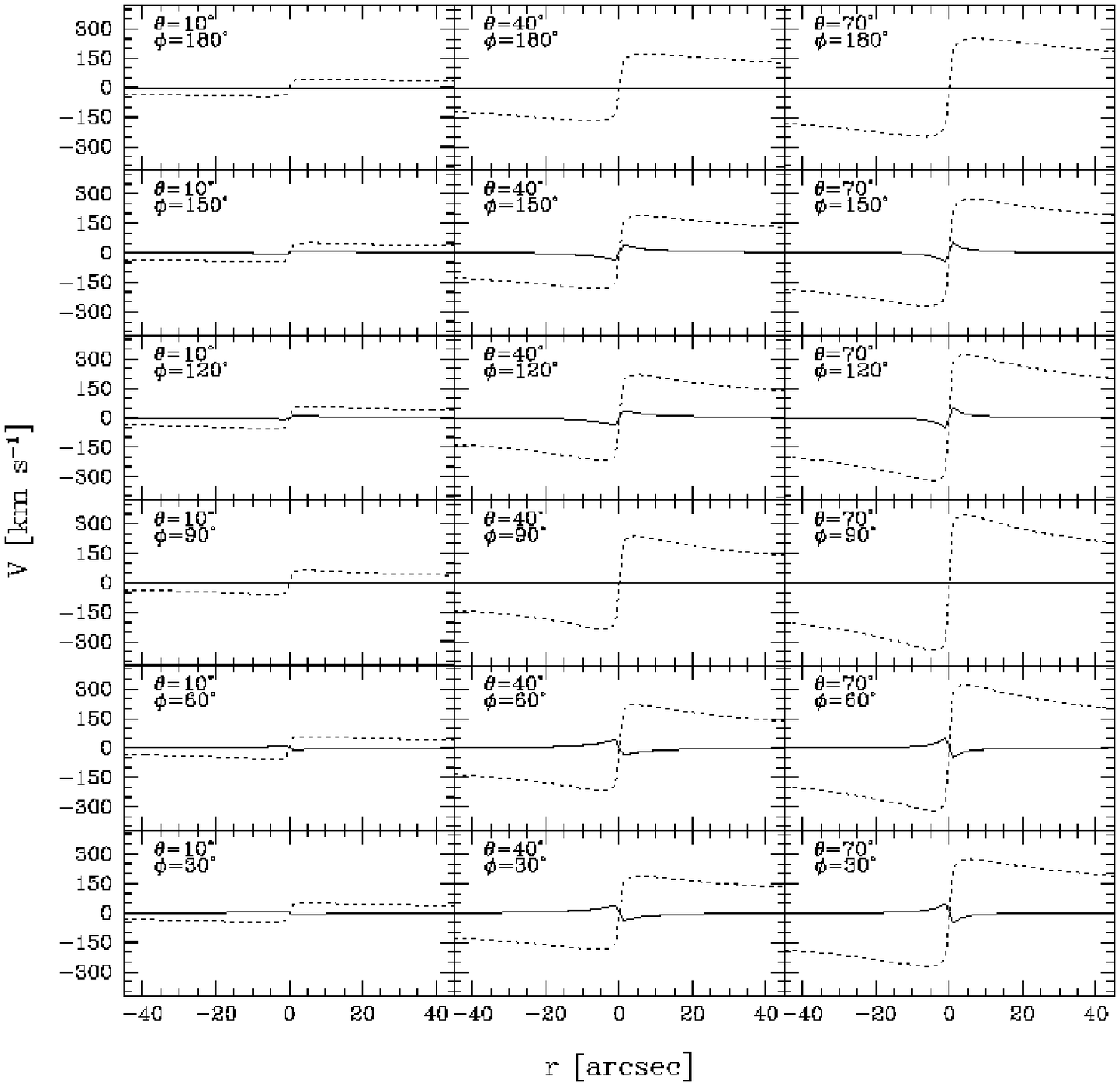}} 
\caption{Ionized-gas kinematics along the galaxy major ({\it dotted
  line\/}) and minor ({\it continuous line\/}) axis moving onto the
  plane perpendicular to the short axis of a triaxial bulge with axial
  ratios $p = 0.7$ and $q = 0.6$, total mass $M=7 \cdot 10^{10}$
  M$_\odot$, scalelength $r_e = 1.5$ kpc, and density profile
  cuspiness $\gamma=1.5$ at a distance $D=25$ Mpc. The viewing angles
  $\theta$ and $\phi$ are specified in each panel.}
\label{fig:atlas}
\end{figure*}

The presence of a gas velocity gradient along both the major and minor
axis allows us to disentangle the case of gas non-circular motion from
that of an inner gaseous disk in orthogonal rotation with respect to
the galaxy disk. Indeed, as discussed in Sect. \ref{sec:ipd}, the
velocity field of a disk galaxy hosting in its center an
orthogonally-rotating gaseous component is characterized by the
presence of both a velocity gradient along the galaxy minor axis and a
zero-velocity plateau along its major axis.
To illustrate this point, in Fig. \ref{fig:atlas} we show the
ionized-gas rotation curves along the galaxy major and minor axis in a
triaxial bulge viewed from various directions.
We considered the gas motion in the family of non-rotating triaxial
mass models with a central density cusp studied by de Zeeuw \& Carollo
(1996). They are a generalization of the spherical $\gamma-$models
with density profile
\begin{eqnarray*}
\rho(r) = \frac{(3-\gamma)}{4 \pi} \frac{M r_0}{r^{\gamma} (r+r_0)^{4-\gamma}},
\end{eqnarray*}
(see Dehnen 1993 and references therein), where $\gamma$ is the
cuspiness of the central density profile ($0 \leq \gamma < 3$), $M$ is
the total mass of the galaxy, and $r_0$ is a scalelength which is
related to the galaxy effective radius $r_e$.  The spherical models
are made triaxial by adding two low-order spherical harmonic terms to
the potential.  The intrinsic shape of the mass density depends on
$(p_0, q_0)$ and $(p_\infty, q_\infty)$, which are the axial ratios of
the triaxial surfaces of constant density at small and at large radii,
respectively. In the absence of figure rotation, settled gas is
rotating either onto the plane perpendicular to the short axis (if $1
> p > q > 0$), or onto the plane perpendicular to the long axis (if $q
> 1 > p > 0$). For large ranges of flattenings and cusp slopes, the
gas velocity field in the above potentials are accurately described by
a first-order epicyclic treatment (de Zeeuw \& Carollo 1996). This
approach has been used by Pizzella et al. (1997) to model the observed
velocity field of the ionized-gas disks of a number of elliptical
galaxies.

We took into account a triaxial bulge with the short axis
perpendicular to the stellar disk. We assumed the axial ratios to be
the same at small and large radii ($p \equiv p_0 = p_\infty$, $q
\equiv q_0 = q_\infty$) and we adopted $p = 0.7$ and $q = 0.6$, which
are the values observed for the Sa NGC 4845 (Bertola et al.  1986;
Gerhard et al. 1986). We assumed $M = 7 \cdot 10^{10}$ M$_\odot$ and
$r_e = 1.5$ kpc for the bulge total mass and effective radius,
respectively. These are typical values for early-type bulges as
derived from dynamical models based on stellar dynamics (Fillmore,
Boroson \& Dressler 1986; Cinzano et al.  1999; Corsini et al.  1999;
Pignatelli et al.  2001) and from surface photometry assuming a de
Vaucouleurs profile (Kent 1985), respectively.
For the model we assumed $D=25$ Mpc, which is the mean distance of the
sample galaxies in which the minor-axis rotation of the gas is
confined to the bulge-dominated region. This distance corresponds to
a scale of 121 pc arcsec$^{-1}$. Finally, we fixed $\gamma=1.5$ to
approximate the bulge surface-brightness profile with a de Vaucouleurs
law.
We derived the gas velocity field at different viewing angles
$(\theta, \phi)$ which are defined as the standard spherical
coordinates of the line of sight in the bulge coordinate system. For
each set of $(\theta, \phi)$ we extracted the gas rotation curve along
the apparent major and minor axis of the galaxy disk (Fig.
\ref{fig:atlas}). It is evident from this figure that the non-circular
gas motions induced by the bulge triaxiality give rise to non-zero gas
velocities observed in the innermost regions of the galaxy minor
axis. The amplitude and radial extension of this phenomenon are
comparable to those we measured in NGC 3885, NGC 4224, and NGC 4586 if
an intermediate inclination is considered. None of the simulated gas
velocity fields shows either a zero-velocity plateau or a change in
the slope of the velocity gradient along the galaxy major axis similar
to those observed in NGC 2855 and NGC 7049.

\subsection{The case of inner polar gaseous disks}
\label{sec:ipd}

The gas velocity field of NGC 2855 and NGC 7049 differs from that of
NGC 3885, NGC 4224, and NGC 4586 since the velocity gradient observed
along the minor axis is associated with a change in the slope of the
velocity gradient measured along the major axis (which is shallower in
the center and steeper away from the nucleus).
This is the kinematic signature of the presence of two decoupled
gaseous components rotating around two roughly orthogonal axes. Taking
into account the triaxiality of bulges, these two
orthogonally-decoupled components have to be settled onto the two
principal planes of the host bulge. The outer gaseous component lies
on the plane perpendicular to the short axis of the bulge, which is the
one containing the galaxy disk. The inner gaseous component rotates in
the plane perpendicular to the long axis of the bulge, and therefore it is
in orthogonal rotation with respect to the galaxy disk. For this
reason we observe in the bulge-dominated region a zero-velocity
plateau along the disk major axis (or at least a shallower velocity
gradient depending on the amount of decoupled gas and spatial
resolution of the kinematic data) and a velocity gradient along the
disk minor axis. The innermost gas component can be identified with an
inner polar gaseous disk.

In both galaxies we exclude that the velocity gradient measured along
the minor axis is due to non-circular gas motion in a barred
potential. We do not observe any bar structure in the low-inclined
disk of NGC 2855 neither on the optical image by Corsini et al.
(2002) nor in the near-infrared images by Peletier et al. (1999) and
M\"ollenhoff \& Heidt (2001). The same is true for NGC 7049 according
to the color map by Veron-Cetty \& Veron (1988). 
In spite of looking more like to an elliptical than any of the other
sample galaxies, NGC 7049 is a disk galaxy as it results from the
photometric decomposition of the $r$-band surface photometry by
Bertola, Vietri \& Zeilinger (1990, 1991).

\addtocounter{table}{2}
\begin{table*}[!ht]
\caption{Galaxies with inner polar disks} 
\begin{tabular}{lllrccrccc}
\hline 
\noalign{\smallskip} 
\multicolumn{1}{c}{Name} & 
\multicolumn{2}{c}{Morphological type} &
\multicolumn{1}{c}{$D$} & 
\multicolumn{1}{c}{$i$} &  
\multicolumn{1}{c}{Inner polar component} & 
\multicolumn{1}{c}{$R$} &
\multicolumn{1}{c}{$d/D_{25}$} & 
\multicolumn{1}{c}{Nuclear bar} & 
\multicolumn{1}{c}{Ref.} \\
\noalign{\smallskip}
\multicolumn{1}{c}{} &  
\multicolumn{1}{c}{[RSA]} & 
\multicolumn{1}{c}{[RC3]} & 
\multicolumn{1}{c}{[Mpc]} &
\multicolumn{1}{c}{[$^\circ$]} &
\multicolumn{1}{c}{} & 
\multicolumn{1}{c}{[pc]} &
\multicolumn{1}{c}{} &  
\multicolumn{1}{c}{} & 
\multicolumn{1}{c}{} \\
\noalign{\smallskip}
\multicolumn{1}{c}{(1)} &
\multicolumn{1}{c}{(2)} & 
\multicolumn{1}{c}{(3)} & 
\multicolumn{1}{c}{(4)} & 
\multicolumn{1}{c}{(5)} & 
\multicolumn{1}{c}{(6)} & 
\multicolumn{1}{c}{(7)} & 
\multicolumn{1}{c}{(8)} &
\multicolumn{1}{c}{(9)} &
\multicolumn{1}{c}{(10)} \\
\noalign{\smallskip} 
\hline 
\noalign{\smallskip} 
Arp 220  &               & S?          & 72.1 & 77 & gaseous disk         & 100 & 0.01 & no & 1\\
NGC 253  & Sc(c)         & SABc(s)     &  3.4 & 79 & gaseous ring         & 150 & 0.01 &    & 2\\ *
NGC 2217 & SBa(s)       &(R)SB0$^+$(rs)& 19.1 & 22 & gaseous disk         & 190 & 0.01 & yes& 3,4\\
NGC 2681 & Sa            & (R')SABa(rs)& 13.3 & 25 & gaseous/stellar disk & 130 & 0.02 & ?  & 5,6,7\\ *
NGC 2841 & Sb            & Sb(r):      &  9.8 & 64 & gaseous disk         & 190 & 0.02 & no & 8,9\\ *
NGC 2855 & Sa(r)         & (R)S0/a(rs) & 22.3 & 27 & gaseous disk         & 220 & 0.03 & no & 10,11\\
NGC 3368 & Sab(a)        & SABab(rs)   &  9.8 & 47 & gaseous disk         & 100 & 0.01 & yes& 7,12\\ *
NGC 4548 & SBb(rs)       & SBb(rs)     & 17.0 & 37 & gaseous disk         & 250 & 0.02 & no & 13\\ 
NGC 4672 &               & Sa(s) pec sp& 39.9 & 75 & stellar disk         & 580 & 0.05 &    & 14,15\\
NGC 4698 & Sa            & Sab(s)      & 17.0 & 65 & gaseous/stellar disk & 250 & 0.03 & no & 15,16,17\\
NGC 5850 & SBb(rs)       & SBb(r)      & 32.4 & 30 & gaseous disk         & 630 & 0.03 & ?  & 5,7,18\\ *
NGC 6340 & Sa(r)         & S0/a(s)     & 19.8 & 26 & gaseous disk         & 480 & 0.05 & no & 19\\
NGC 7049 & S0$_{3}$(4)/Sa& S0(s)       & 29.6 & 47 & gaseous disk         & 430 & 0.02 & no & 11\\
NGC 7217 & Sb(r)         & (R)Sab(r)   & 16.4 & 35 & gaseous disk         & 240 & 0.03 & no & 20,21\\
NGC 7280 &               & SAB0$^+$    & 26.3 & 52 & gaseous disk         & 260 & 0.03 & no & 22\\
\noalign{\smallskip} 
IC 1689  &               & S0(s)       & 63.0 & 55 & gaseous/stellar ring &2140 & 0.24 &    & 23,24\\ *
UGC 5600 &               & S0?         & 37.6 & 46 & gaseous ring         &1090 & 0.14 &    & 25,26\\ * 

\noalign{\smallskip}  
\hline 
\noalign{\smallskip}  
\end{tabular}\\  
\begin{footnotesize}
\begin{minipage}{18cm} 
NOTES. -- 
Col. 2: Morphological classification from RSA.
Col. 3: Morphological classification from RC3.
Col. 4: Distance either from the listed references or from
        Col. 20 in RSA assuming $H_0=75$ \kmsmpc .
Col. 5: Inclination of the galaxy disk either from listed references or
        from the observed axial ratio $R_{25}$ (RC3) after correcting
        for intrinsic axial ratio as in Guthrie (1992).          
Col. 6: Properties of the inner polar component.
Col. 7: Radius of the inner polar component corresponding to the
        maximum observed velocity.
Col. 8: Ratio of the inner polar component to galaxy size. 
        Galaxy isophotal diameters are from RC3. 
Col. 9: Presence of a nuclear bar. ? = uncertain.
Col. 10: References for kinematic and photometric data. 
 1 = Eckart \& Downes (2001);
 2 = Anantharamaiah \& Goss (1996);
 3 = Bettoni, Fasano \& Galletta (1990);
 4 = Jungwiert, Combes \& Axon (1997);
 5 = Wozniak et al. (1995);
 6 = Erwin \& Sparke (1999);
 7 = Moiseev, Vald\'es \& Chavushyan (2003);
 8 = Sil'chenko, Vlasyuk \& Burenkov (1997);
 9 = Afanasev \& Sil'chenko (1999);
10 = Corsini, Pizzella \& Bertola (2002);
11 = this paper;
12 = Sil'chenko et al. (2003);
13 = Sil'chenko (2002);
14 = Sarzi et al. (2000);
15 = Bertola \& Corsini (2000);
16 = Bertola et al. (1999);
17 = Pizzella et al. (2002);
18 = Buta \& Croker (1993); 
19 = Sil'chenko (2000);
20 = Zasov \& Sil'chenko (1997);
21 = Sil'chenko \& Afanasev (2000);
22 = Afanasev \& Sil'chenko (2000);
23 = Reshetnikov, Hagen-Thorn \& Yakovleva (1995);
24 = Hagen-Thorn \& Reshetnikov (1997);
25 = Karataeva et al. (2001);
26 = Shalyapina, Moiseev \& Yakovleva (2002).
\end{minipage} 
\end{footnotesize}
\label{tab:ipd}
\end{table*}

Galaxies hosting an inner polar disk are a new class of objects, since
these orthogonally-decoupled structures have been discovered in the
last few years. The investigation of their structural properties and
formation processes offers some clues about the processes driving
secular evolution of gaseous and stellar components in galaxy centers.
In order to get an exhaustive picture of the phenomenon, we compiled a
list of disk galaxies with an inner polar disk from both data of the
present paper and the literature. The main properties of both the host
galaxies and their inner polar components are given in Table
\ref{tab:ipd}.  For comparison we also included in Table \ref{tab:ipd}
IC 1689 and UGC 5900, which are characterized by a small polar ring
rather than an inner polar disk.

Inner polar disks are disks of small size ($R\ap300$ pc), which are
located in the center of lenticular and spiral galaxies and are
rotating in a plane perpendicular to that of the main disk of their
host. Inner polar gaseous disks have been reported in all the galaxies
listed in Table \ref{tab:ipd}, except for NGC 4672 (Sarzi et al. 2002)
where the orthogonally-rotating component is made up only of
stars. Gas and stars are observed in the inner polar disks of NGC 2681
(Moiseev et al. 2003) and NGC 4698 (Bertola et al. 1999; Bertola \&
Corsini 2000; Pizzella et al. 2002).

The radial extent of inner polar disks is small with respect to that
of their host galaxy ($d/D_{25}\ap0.03$). They are smaller than the
so-called `inner' polar rings found in IC 1689 ($d/D_{25}=0.24$,
Reshetnikov, Hagen-Thorn \& Yakovleva 1995) and UGC 5900
($d/D_{25}=0.14$, Karataeva et al.  2001). On the other hand, the
diameter of `classical' polar rings is comparable or even larger than
that of the galaxies they surround ($d/D_{25}\ap1-3$, see Sackett 1991).

According to morphological classification and isophotal analysis of
their host galaxies, the presence of an inner polar disk is not
directly related either to the presence of a main bar or to that of
a nuclear bar. NGC 2217, NGC 4845 and NGC 5850 are the only bona-fide
barred spirals of the sample in Table \ref{tab:ipd}. An
intermediate-scale bar has been recognized in NGC 2841 by Afanasev \&
Sil'chenko (1999).  Nuclear bars have been photometrically detected in
NGC 2217 (Jungwiert et al. 1997), NGC 2681 (Wozniak et al. 1995; Erwin
\& Sparke 1999), NGC 3368 (Jungwiert et al. 1997), and NGC 5850 (Buta
\& Crocker 1993; Wozniak et al. 1995), although only the nuclear bar
of NGC 3368 has been kinematically confirmed by Moiseev et al. (2003)
by means of integral-field spectroscopy.

The short dynamical time (few Myr) we derived for the galaxy regions
where we observe the orthogonal rotation implies that inner polar
disks are already settled in an equilibrium configuration.
We exclude the possibility taken into account by Sil'chenko (2002)
that gas of inner polar disks is moving on anomalous orbits in a
triaxial bulge (or bar) that is tumbling about its short axis. In
fact we do not observe gas in retrograde motion relative to stars at
large radii from the galaxy center, where the gas orbits are expected
to be highly inclined with respect to the figure rotation axis (see
van Albada, Kotanyi \& Schwarzschild 1982; Friedli \& Benz 1993 for
details).
We therefore suggest that inner polar disks lie in the principal plane
perpendicular to the long axis of the triaxial bulge (or of the bar
when it is present). 

As for `classical' polar rings (see the recent numerical results by
Bournaud \& Combes 2003), the acquisition of external gas via merging
or accretion on nearly polar orbits by a pre-existing galaxy has been
usually proposed to account for the formation of
kinematically-decoupled components like inner polar disks (see list of
references in Table \ref{tab:ipd}). In some objects this scenario is
supported by different arguments. The S0/a NGC 2681 underwent some
kind of widely distributed starburst not earlier than 1 Gyr ago, which
has been ascribed to the dumping of tidally extruded gas from a galaxy
neighbor (Cappellari et al. 2001).  The Sa NGC 2855 is surrounded by a
ring-like structure elongated in a direction close to the galaxy minor
axis, which is indicative of a second event (Corsini et al. 2001).
The same is true for the Sab NGC 3368 (Sil'chenko et al. 2003) which
is close to the supergiant intergalactic ring of neutral hydrogen
found in the Leo I group by Schneider et al. (1983). The large-scale
distribution of the neutral hydrogen of the SBb NGC 5850 has been
explained as due to an high-speed encounter with its companion NGC
5846 (Higdon, Buta \& Purcell 1998). The Sab NGC 7217 hosts a
counterrotating stellar disk, which is suggestive of a retrograde
accretion from the environment (Kuijken \& Merrifield 1994).

Sil'chenko (2001) argued that the inner polar disks of NGC 2841, NGC
4548, NGC 6340 and NGC 7217 have an internal origin as their hosts are
isolated spirals with a large amount of normally-rotating gas, and
invoking the qualitative scenario proposed by Sofue \& Wakamatsu
(1994) to put the gas on the polar orbits in the center of barred
galaxies. However, all these arguments do not exclude an external
origin of the decoupled component. Bettoni, Galletta \& Prada (2001)
proved that the environment of galaxies that experienced past gas
accretion do not appear statistically different from those of normal
galaxies. This means that second events are not ruled out by the
isolation of the host galaxy.  A large amount of accreted gas can sink
toward the galaxy nucleus, sweeping away the pre-existing gas and
giving origin to the decoupled component. This kind of interaction
between newly acquired and pre-existing gas has been proposed to
account for the reversal of rotation measured in the gaseous disk of
NGC 4826 (Rubin 1994). Finally, the dynamical process investigated by
Sofue \& Wakamatsu (1994) does not prevent the gas driven to the
galaxy center from having been acquired.

To address the frequency of inner polar disks, a kinematic survey of a
complete sample of nearby bulges is highly desirable. Indeed their
small size prevents statistics based on morphological classification
by visual inspection of photographic plates as done for classical
polar rings (Whitmore et al. 1990). The measurement of the gas
velocity field at high spatial resolution in the center of disk
galaxies by state-of-the-art integral-field spectrographs (e.g.  {\tt
SAURON} and {\tt MPFS}, see the volume edited by Rosado et al. 2003
for a review) represents the ideal tool for this kind of
investigation.

\acknowledgements 

We are indebted to R. Bender and R. Saglia for providing us with the
FCQ package we used for measuring the stellar kinematics.
We thank A. Moiseev for helpful discussions, and O. Sil'chenko for
making available her data in advance of publication.
This research has made use of the Lyon-Meudon Extragalactic Database
(LEDA) supplied by the LEDA team at CRAL-Observatoire de Lyon (France)
and Digitized Sky Survey (DSS) produced at the Space Telescope Science
Institute under U.S. Government grant NAG W-2166.

\end{document}